\documentstyle[12pt,epsfig]{article}
\newcommand{\be}{\begin{equation}}
\newcommand{\ee}{\end{equation}}
\newcommand{\bea}{\begin{eqnarray}}
\newcommand{\eea}{\end{eqnarray}}

\newcommand{\ggam}{\gamma\gamma}
\newcommand{\cbc}{c\bar c}

\newcommand{\beq}{\begin{equation}}
\newcommand{\eeq}{\end{equation}}

\def\ga{\mathrel{\mathpalette\fun >}}
\def\fun#1#2{\lower3.6pt\vbox{\baselineskip0pt\lineskip.9pt
\ialign{$\mathsurround=0pt#1\hfil##\hfil$\crcr#2\crcr\sim\crcr}}}

\begin{document}

\title{Charmed quark component of the photon wave function}

\author{V.V. Anisovich, L.G. Dakhno,
V.N. Markov, V.A. Nikonov and A.V.  Sarantsev}
\date{27.10.2004}
\maketitle

\begin{abstract}

 We determine the $c\bar c$ component of the photon wave function on
 the basis of \\(i) the data on the transitions
$e^+e^- \to J/\psi(3096), \psi(3686),\psi(4040),\psi(4415)$,
\\(ii) partial widths of the
 two-photon decays $\eta_{c0}(2979),\chi_{c0}(3415),
\chi_{c2}(3556)\to
 \gamma\gamma$, and \\(iii) wave functions of the charmonium states
 obtained by solving the Bethe-Salpeter equation for the $c\bar c$
 system. Using the  obtained $c\bar c $ component of the photon wave
 function we calculate the $\gamma\gamma$ decay partial widths for
 radial excitation $2S$ state, $\eta_{c0}(3594)\to \gamma\gamma$,
 and $2P$ states $\chi_{c0}(3849), \chi_{c2}(3950)\to \gamma\gamma$.

\end{abstract}

\section{Introduction}

There is a number of processes that can be sensibly treated by
introducing  the $c\bar c$ component of the photon wave function.
First, it is the production of charmonium in the two-photon
transitions such as $\gamma^*\gamma^*\to c\bar c$-mesons, production of
$\psi$-mesons in the $e^+e^-$-annihilation and production of charmonia
in photon-nucleon collisions, $\gamma^*p\to c\bar c$-${\rm meson}+X$.
In the present paper, we determine the $c\bar c$-component of photon
wave function, or the transition vertex $\gamma\to c\bar c$, following
the method developed in [1,2], where quark--antiquark components were
found for the transitions $\gamma\to u\bar u, d\bar d, s\bar s$.

The method of introducing quark--antiquark photon wave function may be
clearly illustrated by considering the
$c\bar c$-${\rm meson}\to \gamma\gamma$ decay.
Dealing with the time-ordered processes, that is
necessary in the dispersion relation approach or light-cone variable
technique, the
$c\bar c$-${\rm meson}\to \gamma\gamma$ decay is a two-step process:
first, the emission of photon by quark (Fig. 1a) or antiquark (Fig. 1b)
and, second, subsequent annihilation $c\bar c\to \gamma$.

In Refs. \cite{1,2}, the triangle diagrams of Figs. 1a,b were treated in
terms of double dispersion relation representation.
The double spectral integral
cuttings of the diagram in Fig. 1a are shown on
Fig. 1c. In the diagram
of Fig. 1c, on the left from the first cutting,  there
is the transition vertex of charmonium,  $G_{charmonium}(s)$,
decaying into $c\bar c$ pair,  where $s$ is the quark invariant
energy squared.
 In light-cone
variables
\be
 \label{1}
s=\frac{m^2_c+k^2_\perp}{x(1-x)}\ ,
\ee
where $m_c$ is the mass of $c$-quark and $(x,{\bf k}_\perp)$ are the
characteristics of one of the quarks (fraction of the momentum along the
$z$-axis and transverse component). In the dispersion integral, the
left-hand cutting leads to the factor
 $G_{charmonium}/(s-M^2_{charmonium})$,
where $M_{charmonium}$ is the charmonium mass, this factor being the
wave function of the initial $c\bar c$-meson:
\be
 \label{2}
\frac{G_{charmonium}(s)}{s-M^2_{charmonium}}=\Psi_{charmonium}(s)\ .
\ee
Likewise, the right-hand cut in Fig. 1c, by describing the transition
$c\bar c\to \gamma$, gives us the factor:
\be
 \label{3}
\frac1{s'}\, e_c\ ,
\ee
where $s'$ is the invariant energy square of quarks in the final state
and $e_c$ is the $c$-quark charge.

When we deal with the
transition $c\bar c\to \gamma$, the interaction of quarks should be
specially taken into consideration.
 The quarks may interact both in the initial (Fig. 2a) and final (Fig.
2b) states. In fact, the interaction of quarks in the initial state
has been accounted for by Eq. (\ref{2}), because the vertex function
$G_{charmonium}$ (or wave function $\Psi_{charmonium}$)
is the solution of
Bethe-Salpeter equation --- diagrammatically, this equation is shown in
Fig. 3a. As to quark interaction in the final state, it should be
particularly taken into account in addition to the point-like
interaction (\ref{3}).
The diagram shown in Fig. 3b stands for the description of quark
interaction in the transition $c\bar c\to \gamma$, and we approximate
it with the sum of the $\psi$-mesons pole terms, see Fig. 3c.
Accordingly, the factor related to the right-hand cut of Fig. 1c is
written as follows:
\be
\label{4}
\frac{G_{\gamma\to c\bar c}(s')}{s'}\, e_c\ ,
\ee
where the vertex function $G_{\gamma\to c\bar c}(s')$ at $s'\sim 4m_c^2$
is a superposition of vertices  of the $S$-wave $\psi$-mesons (see
Fig. 3c):
\be
\label{5}
G_{\gamma\to c\bar c}(s)\simeq \sum_n C_n G_{\psi(nS)}(s)\ , \quad
s\sim 4m_c^2\ .
\ee
Here $n$ is the radial quantum number of $\psi$-meson and $C_n$'s are
numerical
 coefficients --- their definition is the task of this paper.
At large $s$, the vertex $c\bar c\to \gamma$ is a point-like one:
\be
 \label{6}
G_{\gamma\to c\bar c}(s)\simeq 1 \qquad {\rm at}\quad s>s_0\ .
\ee
The parameter $s_0$ can be determined using the data on
$e^+e^-$-annihilation into hadrons: it is defined by the energy range,
where the ratio $R(s)=\sigma(e^+e^- \to {\rm hadrons})/ \sigma(e^+e^-
\to \mu^+\mu^-)$ reaches a constant-behaviour regime above the
threshould of the charm production, $R(s)\simeq 10/3$. The data support
the value $s_0 \sim (10-15)$ GeV$^2$  \cite{R}.

Therefore, to describe the transition $c\bar c \to \gamma$ we may
introduce a characteristics, which, similarly to (\ref{2}), may be
called
the charmed quark component of photon wave function:
\be
 \label{7}
\frac{G_{\gamma\to c\bar c}(s)}{s-q^2}=\Psi_{\gamma(q^2)\to c\bar
c}(s)\ ,
\ee
here $q$ is the photon four-momentum. Let us emphasize that such a wave
function is determined at $s\ga 4m^2_c$.

There exists a reaction which is directly related to the
photon wave function. This is the transition $e^+e^- \to \psi(nS)$, see
Fig. 4: here the loop diagram is defined by the convolution of
meson wave function and the vertex $\gamma\to c\bar c$:
$\Psi_{nS}\otimes G_{\gamma\to c\bar c}$.

Dealing with the $c\bar c$ interaction in the transition
$\gamma\to c\bar c$, we take into consideration the $S$-wave
$\psi$-mesons only, while  the contribution of
mesons dominated by the $D$-wave such as $\psi(3770)$ and
$\psi(4160)$, is neglected. The error coming from such a
neglect can be evaluated by comparing the $\psi$-meson production
cross sections for $S$- and $D$-wave states
 in the $e^+e^-$-annihilation: it is of the order of 10\%.
The $D$-wave admixture into the low-lying $1^{--}$-mesons
is also small: it is of the order of 1\% for $J/\psi(1S)$ and $\psi(2S)$
\cite{4}.

We are performing calculations of two- and three-point loop diagram in
the spectral integration technique. All the equations used, up to
trivial substitutions of quark masses and charges, were obtained in
\cite{BS}. Because of that, in this paper we present final
expressions only accompanied by necessary comments and references.

The main difference between calculations  for $\gamma\to
c\bar c$ and those for $\gamma\to u\bar u,d\bar d, s\bar s$
carried out in \cite{1,2}
consists in our regard towards quark wave functions. In
\cite{1,2}, we used phenomenological quark wave functions for $\pi^0,
\eta,\eta'$, while for light vector mesons ($\rho,\omega,\phi$) we
assumed the quark wave functions to be similar to analogous
pseudoscalar-meson wave functions, with whom they form the lowest
36-plet in terms of SU(6)-symmetry. But here, when
reconstructing the wave function for the transition $\gamma\to c\bar
c$, we have applied the charmonium wave functions found out from  the
solution of Bethe-Salpeter equation \cite{4}.

It was  a long history of calculation of charmonium states, and now
there is a rich collection of results obtained in the framework of the
nonrelativistic approaches \cite{Hulth,Godfrey,Gupta,Lucha} as well
as in different variants of the Bethe-Salpeter equation
\cite{Linde,Resag,T-Tjon,Hersbach},
see also references therein. However, one should bear in mind that in
calculations related to quark--antiquark systems, one
focuses as a rule on the description of levels (or masses) of the system.
Had the potential or its relativistic analogue been known, the
Bethe--Salpeter equation would undoubtedly provide us with both levels
and wave functions. But the problem is that in fact our  knowledge about
the quark--quark interaction in the soft region is rather poor.
Therefore, in the reconstruction of quark--antiquark levels one can
obtain a variety of wave functions. Within the spectral
integration technique used here, we see that unambigous
determination of interaction is possible in the simultaneous
description of both  levels and wave functions, see discussion
in Section 2.6.3 of \cite{BS}.
Because of that,  by describing  the $c\bar c$ system, we
have used in \cite{4} as an
input both the known levels and known values of radiative transitions
$(c\bar c)_{in}\to \gamma+(c\bar c)_{out}$. These radiative transitions
are rather sensitive to wave functions, thence a comparatively good
description of radiation transitions in \cite{4} allows us to believe
that the wave functions of lowest $c\bar c$ states are determined
reliably too.

The paper is organized as follows. In Section 2 we present briefly
the information about the spectral-integration Bethe-Salpeter
equation, in the framework of
which the description of the $c\bar c$ systems was performed in
\cite{4}. In Section 2 of the present article we also give the
charmonium wave functions obtained at the determination of
$G_{\gamma\to c\bar c}$. In Section 3 we write down the formulae for
the transition amplitudes $e^+e^- \to \psi$ and $\eta_{c0},\chi_{c0},
\chi_{c2}\to \gamma\gamma$, which are used in the fit, and present the
results for the photon wave function. Brief summary is given in
Conclusion.

\section{Charmonium wave functions}

The spectral integral equation  for quark-antiquark wave function,
which can be conventionally called as the Bethe-Salpeter equation,
is written for a system with
the total momentum $J$,
angular momentum $L$,
quark--antiquark spin $S$ and radial number $n$.
For the $c \bar c$ system it reads:
\be
\left(s-M_{charmonium}^2\right)
\widehat\Psi^{(S,L,J)}_{(n)\,\mu_{1}\cdots\mu_{J}} (k_{\perp})=
\label{bs1}
\ee
$$
=\int \limits_{4m_c^2}^{\infty}\frac{ds'}{\pi} d\Phi_2(P';k'_1 ,k'_2)\,
\widehat V\left(s,s',(k_{\perp}k'_{\perp})\right)(\hat k_1'+m_c)
\widehat \Psi^{(S,L,J)}_{(n)\,\mu_{1}
\cdots\mu_{J}} (k'_{\perp})(-\hat k_2'+m_c)\ .
$$
Here the
quarks are mass-on-shell, $k_1^2=k_1'^2=k_2^2=k_2'^2=m_c^2$,
and the phase space factor in the intermediate state is determined
as follows:
\be
d\Phi_2(P';k'_1 ,k'_2) = \frac 12 \frac{d^3k'_1}{(2\pi)^3\, 2k'_{10}}
\frac{d^3k'_2}{(2\pi)^3\, 2k'_{20}}
(2\pi)^{4}\delta^{(4)}(P'-k'_1 -k'_2)\ .
\ee
We use the following notations:
\be
k_\perp=\frac12\left(k_1-k_2\right)\ , \quad
P=k_1+k_2\ ,
\qquad k'_\perp=\frac12\left(k'_1-k'_2\right)\ , \quad
P'=k'_1+k'_2\ ,
\label{bs2}
\ee
$$
P^2=s\  ,\quad \ P'^2=s'\  , \qquad
g^{\perp}_{\mu\nu} =g_{\mu\nu}-
\frac{P_\mu P_\nu}{s} \  , \quad
g'^{\perp}_{\mu\nu} =g_{\mu\nu}-
\frac{P'_\mu P'_\nu}{s'} \  ,
$$
so one can write
$\ k^{\perp}_{\mu}=k_{\nu}g^{\perp}_{\nu\mu}$ and
$\ k'^{\perp}_{\mu}=k'_{\nu}g'^{\perp}_{\nu\mu}$\ .
In the centre-of-mass system, the integration  can be
re-written as
\be
\int \limits_{4m_c^2}^{\infty}\frac{ds'}{\pi} d\Phi_2(P';k'_1
,k'_2)\longrightarrow \int \frac{d^3k'}{(2\pi)^3k'_0}\ ,
\ee
where $k'$ is the momentum of one of the quarks.
The wave function reads:
\be
\widehat
\Psi^{(S,L,J)}_{(n)\,\mu_{1}\cdots\mu_{J}}(k_{\perp})=
Q^{(S,L,J)}_{\mu_{1}\cdots\mu_{J}}(k_{\perp})\,
\psi^{(S,L,J)}_n (k_{\perp}^2)\ .
\label{bs3}
\ee
Here $Q^{(S,L,J)}_{\mu_{1}\cdots\mu_{J}}(k_{\perp})$    is
the moment operator for fermion-antifermion
system \cite{operators} defined as follows:
\bea
\label{bs10}
&& Q^{(0,J,J)}_{\mu_1\mu_2\ldots\mu_J}(k_{\perp})=i\gamma_5
X^{(J)}_{\mu_1\ldots\mu_J}(k_{\perp})\ ,
\\
&& Q^{(1,J+1,J)}_{\mu_1\ldots\mu_J}(k_{\perp})=\gamma^{\perp}_\alpha
X^{(J+1)}_{\mu_1\ldots\mu_{J}\alpha}(k_{\perp})\ ,
\nonumber
\\
&& Q^{(1,J,J)}_{\mu_1\ldots\mu_J}(k_{\perp})=\frac{1}{\sqrt s} \,
\varepsilon_{\alpha\nu_1\nu_2\nu_3}\gamma^{\perp}_{\alpha}P_{\nu_1}
Z^{(J)}_{\nu_2\mu_1\ldots\mu_J,\nu_3} (k_{\perp})\ ,
\nonumber
\\
&& Q^{(1,J-1,J)}_{\mu_1\ldots\mu_J}(k_{\perp})=
\gamma^{\perp}_\alpha
Z^{(J-1)}_{\mu_1\ldots\mu_J,\alpha}(k_{\perp})\ ,
\nonumber
\eea
where
\bea
&&X^{(J)}_{\mu_1\ldots\mu_J}(k_\perp)
=\frac{(2J-1)!!}{J!}
\bigg [
k^\perp_{\mu_1}k^\perp_{\mu_2}k^\perp_{\mu_3}k^\perp_{\mu_4}
\ldots k^\perp_{\mu_J} -
\label{x-direct}
\\
&&-\frac{k^2_\perp}{2J-1}\left(
g^\perp_{\mu_1\mu_2}k^\perp_{\mu_3}k^\perp_{\mu_4}\ldots k^\perp_{\mu_J}
+g^\perp_{\mu_1\mu_3}k^\perp_{\mu_2}k^\perp_{\mu_4}\ldots
k^\perp_{\mu_J} + \ldots \right)+
\nonumber
\\
&&
+\frac{k^4_\perp}{(2J-1)
(2J-3)}\left(
g^\perp_{\mu_1\mu_2}g^\perp_{\mu_3\mu_4}k^\perp_{\mu_5}
k^\perp_{\mu_6}\ldots k^\perp_{\mu_J}+
\right .
\nonumber
\\
&&
\left .
+g^\perp_{\mu_1\mu_2}g^\perp_{\mu_3\mu_5}k^\perp_{\mu_4}
k^\perp_{\mu_6}\ldots k^\perp_{\mu_J}+
\ldots\right)+\ldots\bigg ]\ ,
\nonumber
\eea
\bea
Z^{(J-1)}_{\mu_1\ldots\mu_J, \alpha}(k_\perp)&=&
\frac{2J-1}{L^2}\left (
\sum^J_{i=1}X^{{(J-1)}}_{\mu_1\ldots\mu_{i-1}\mu_{i+1}\ldots\mu_J}
(k_\perp)g^\perp_{\mu_i\alpha} -\right .
\nonumber \\
&&\left . -\frac{2}{2J-1} \sum^J_{i,j=1 \atop i<j}
g^\perp_{\mu_i\mu_j}
X^{{(J-1)}}_{\mu_1\ldots\mu_{i-1}\mu_{i+1}\ldots\mu_{j-1}\mu_{j+1}
\ldots\mu_J\alpha}(k_\perp) \right )\ .
\nonumber
\eea
The interaction block can be expanded in a series in a full set of the
$t$-channel operators:
\be
\widehat V\left(s, s', (k_{\perp}k'_{\perp})\right)=
\sum_{I} V_I\left(s, s', (k_{\perp}k'_{\perp})\right)
\widehat O_I \otimes \widehat O_I\ ,
\label{bs4}
\ee
$$
\widehat O_I
={\rm I}, \;\; \gamma_{\mu},\;\; i\sigma_{\mu\nu},\;\;
i\gamma_{\mu}\gamma_5, \;\; \gamma_5\ .
$$
The Bethe-Salpeter equation  (\ref{bs1}) is written in the
momentum representation, it was solved in \cite{4} in the
momentum representation  as well. Equation  (\ref{bs1}) allows one
to use as an interaction the instantaneous approximation or to take into
account retardation effects.  In the instantaneous approximation  one
has \be
\widehat V\left(s,s',(k_{\perp}k'_{\perp})\right)
\longrightarrow
\widehat V(t_\perp),  \qquad
t_\perp = (k_{1\perp}-k'_{1\perp})_\mu(-k_{2\perp}+k'_{2\perp})_\mu \,
\ee
The retardation effects are taken into account, when  the
momentum transferred squared $t$ in the interaction block depends on
the time components of quark momenta (for more detail
see Section 2.5 of \cite{BS} and the  discussion in
\cite{Hersbach,her2,gross,maung}):
\be \widehat
V\left(s,s',(k_{\perp}k'_{\perp})\right) \longrightarrow \widehat V(t),
\qquad t = (k_{1}-k'_{1})_\mu(-k_{2}+k'_{2})_\mu\ .
\ee
It turned out that for the fit of the $\cbc$ states we need only two
interaction blocks with the following $t$-dependence:
\bea
I_0(t)&=&\frac{8\pi\mu}{(\mu^2-t)^2}\ ,
\nonumber\\
I_1(t)&=&8\pi\left(\frac{4\mu^2}{(\mu^2-t)^3}-\frac{1}{(\mu^2-t)^2}
\right)\ .
\label{i0i1}
\eea
It also occurred that the results of the fit depended weakly on
whether the $t$- or $t_\perp$-dependence was used
in (\ref{i0i1}). Because of that, for the sake of
simplicity, we present below the results obtained in the instantaneous
approximation substituting $t\to t_\perp$ in (\ref{i0i1}).

Traditionally, the interaction of heavy quarks in the instantaneous
approximation is presented in terms of the potential $V(r)$. The
 form of the  potential can be obtained  with the Fourier
transform of (\ref{i0i1}) in the centre-of-mass system.
Thus we have
\bea
t_\perp&=&-(\vec{k}-\vec{k'})^2=-\vec{q}^{\, 2}\ ,
\nonumber\\
I_c^{(N)}(r,\mu)&=&\int{
\frac{d^3q}{(2\pi)^3}\;e^{-i\vec{q}\, \vec{r}}\;I_N(t_\perp)}\ ,
\label{icin}
\eea
that gives
\bea
I_c^{(0)}(r,\mu)=e^{-\mu r}\ ,     \qquad
I_c^{(1)}(r,\mu)=r\;e^{-\mu r}\ .
\eea
The potential used in [4] had the form
\bea
V(r)= a + b\ r + c\ e^{-\mu\ r}\ ,
\eea
where the constant and linear (confinement) terms read:
\bea
a&\to&a\;I_c^{(0)}(r,\mu_{constant}\to 0)\ ,
\nonumber\\
br&\to&b\;I_c^{(1)}(r,\mu_{linear}\to 0)\ .
\eea
The limits $\mu_{constant}\, ,\mu_{linear}\to 0$ mean
that in the fitting procedure
the parameters $\mu_{constant}$ and $\mu_{linear}$ are chosen to be
small enough, of the order of 1--10 MeV.  It ws checked  that the
solution for the considered states ($n\leq 6$)
was stable when changing $\mu_{constant}$ and $\mu_{linear}$
in this interval, 1--10 MeV .

In \cite{4}, charmonium wave functions were
fitted in the following form:
\bea
\label{21}
\Psi_{charmonium}^{(n)}(s)=e^{-\beta k^2}\sum\limits_{i=1}^9 c_i^{(n)}
k^{i-1}\, .
\eea
Recall that $k^2$ is the relative quark momentum squared,
$s=4m_c^2+4k^2$, and $n$ is the radial quantum number;
$\beta=1$ GeV$^{-2}$ for all $\cbc$ states.

Two solutions with two
types of the $t$-channel exchanges were used:
\bea
\nonumber
{\rm Solution \, I}\,&:& \qquad
{\rm I}\otimes {\rm I}\ ,\quad \gamma_{\mu}\otimes \gamma_{\mu}\ , \quad
\gamma_5\otimes \gamma_5\ , \\
{\rm Solution \, II}&:& \qquad
{\rm I}\otimes {\rm I}\ ,\quad \gamma_{\mu}\otimes \gamma_{\mu}\ .
\nonumber
\eea
Table 1 demonstrates the obtained values of the potential parameters
($a$, $b$, $c$, $\mu$) in Solutions I and II. In all solutions
we put
$m_c=1.25$ GeV.

The measured
masses of the $c\bar c$-states and the results of the fit for
$n=1,2,3,4,5,6$ are displayed in Table 2.

\begin{table}
\caption{Parameters of the potential (in GeV units)}
\begin{center}
\begin{tabular}{|c|c|c|c|c|c|}
\hline
Type of interaction ($\widehat O_I \otimes \widehat O_I$)
& Solution&   $a$  &  $b$ & $c$   & $\mu$ \\
\hline
      Scalar $({\rm I} \otimes {\rm I})$
& I       & -1.527 & 0.170& 1.013 & 0.201 \\
             & II      & -1.417 & 0.158& 0.883 & 0.201 \\
\hline
      Vector  $(\gamma_{\mu} \otimes \gamma_{\mu})$
& I       & -1.539 & 0    & 2.133 & 0.401 \\
             & II      & -1.540 & 0    & 2.130 & 0.401 \\
\hline
Pseudoscalar $(\gamma_{5} \otimes \gamma_{5})$
& I       & -3.000 & 0    & 0     & 0.201 \\
             & II      &  0     & 0    & 0     & - \\
\hline
\end{tabular}
\end{center}
\end{table}

\newpage

\begin{table}
\caption{Measured masses and results of the fit for Solutions I and II
(in GeV units). Bold-faced numbers stand for masses which have been
used in fitting procedure \cite{4}. }
\begin{center} \begin{tabular}{|c|c c|c c|c c|c c|c c|} \hline
n&$\psi (nS)$&&$\eta_c(nS)$&&$\chi_{c0}(nP)$&&$\chi_{c1}(nP)$&&$\chi_{c2}
(nP)$&
\\
\hline
 &data&fit&data&fit&data&fit&data&fit&data&fit \\
\hline
1& {\bf 3.096} & 3.1022&{\bf 2.979} & 2.9776&{\bf 3.415} & 3.3933&
{\bf 3.510} & 3.4962&{\bf 3.556} & 3.5676 \\
2& {\bf 3.686} & 3.6737&{\bf 3.594}& 3.6246& & 3.8485&       & 3.9002&
& 3.9495 \\
3&4.040  & 4.0565& & 4.0225&       & 4.1812&   & 4.2514&   & 4.3046 \\
4&4.415  & 4.3960& & 4.3678&       & 4.5569&   & 4.6709&   & 4.8250 \\
5&       & 4.8465& & 4.8233&       & 5.1185&   & 5.1886&   & 5.4758 \\
6&       & 5.4448&       & 5.4242& & 5.7510&   & 5.8404&   & 6.2197 \\
\hline
1&{\bf 3.096} & 3.1023&{\bf 2.979} & 2.9772&{\bf 3.415}  & 3.3958&
{\bf 3.510 }& 3.4979&{\bf 3.556}   & 3.5687 \\
2&{\bf  3.686} & 3.6721&{\bf 3.594}& 3.6238&       & 3.8447&       &
3.8957&      & 3.9434 \\
3&4.040  & 4.0470& & 4.0139&       & 4.1705&    & 4.2406&  & 4.2976 \\
4&4.415  & 4.3801& & 4.3527&       & 4.5488&    & 4.6605&  & 4.8188 \\
5&       & 4.8322& & 4.8090&       & 5.1123&    & 5.1820&  & 5.4669 \\
6&       & 5.4336& & 5.4135&       & 5.7460&    & 5.8379&  & 6.1986 \\
\hline \end{tabular} \end{center} \end{table}

\begin{table}
\caption{Constants $c_i^{(n)}$ (in GeV units) for the wave functions of
  Eq.(\ref{21}) in Solutions I, II}

\begin{center} {\footnotesize
\begin{tabular}{|c|c|c|c|c|c|c|c|c|c|c|}
\hline
$i$ &$J/\psi(1S)$ &$\psi(2S)$ &$\psi(3S)$ &$\psi(4S)$
&$\eta_c(1S)$ &$\eta_c(2S)$ &$\chi_{c0}(1P)$ &$\chi_{c0}(2P)$
&$\chi_{c2}(1P)$ &$\chi_{c2}(2P)$  \\
\hline
1&   8.15&  -18.87&   43.70&   -68.75&  -7.064&  -17.72&   27.17&  131.60&  -45.15& -144.38 \\
2&   3.01&    9.16& -271.18&   682.41& 0.08171&    4.60&  -12.77& -480.42&   59.15&  569.17 \\
3& -48.02&  266.00&  465.83& -2356.65&  20.748&  247.04& -100.08&  539.03&  141.47& -587.98 \\
4&  76.13& -729.46&  -92.87&  3873.67& -14.849& -653.28&  181.19&  -23.55& -432.83& -315.11 \\
5& -49.16&  871.28& -504.13& -3421.29& -22.992&  770.56& -120.56& -442.03&  481.62& 1126.02 \\
6&   9.17& -566.22&  587.76&  1688.66&  41.115& -500.42&   23.01&  410.77& -286.31& -978.20 \\
7&   4.81&  208.18& -286.06&  -452.00& -25.682&  185.63&   11.05& -172.89&   96.19&  413.89 \\
8&  -2.65&  -40.78&   66.18&    57.34&   7.413&  -37.01&   -6.04&   36.23&  -17.21&  -87.74 \\
9& 0.3712&   3.320&  -5.948&   -2.183& -0.8335&   3.094&  0.8269&  -3.081&   1.274&   7.454 \\
\hline
1&   8.14&  -19.41&   45.87&   -69.8&   -7.061&  -18.16&   27.79&  137.91&  -46.32& -151.59 \\
2&   2.86&   12.56& -295.78&   705.5&  0.29980&    7.16&  -15.64& -520.67&   64.72&  624.91 \\
3& -47.50&  260.19&  559.16& -2482.2&   19.754&  243.34&  -95.29&  633.58&  132.65& -753.22 \\
4&  75.58& -729.85& -261.10&  4172.7&  -13.001& -655.51&  177.81& -131.97& -428.51&  -59.52 \\
5& -48.97&  882.00& -338.96& -3793.5&  -24.954&  781.02& -120.21& -377.56&  484.48&  895.03 \\
6&   9.20& -578.49&  494.77&  1949.0&   42.396& -511.22&   23.98&  393.18& -290.87& -851.60 \\
7&   4.76&  214.44& -256.26&  -554.9&  -26.199&  190.89&   10.43& -172.64&   98.47&  372.46 \\
8&  -2.63&  -42.31&  61.177&    78.7&    7.527&  -38.26&   -5.88&   37.11&  -17.73&  -80.30 \\
9&  0.371&   3.468& -5.6097&   -4.01& -0.84433&   3.212&  0.8118&  -3.211&   1.320&   6.891 \\
\hline
\end{tabular}
}
\end{center}
\end{table}

Figure 5 shows  the radiative decay
 transitions, which were included into fitting
procedure in \cite{4}. One can see the calculated numbers for
partial widths and  experimental values with errors used in
the fit: the 20\%-accuracy was accepted for the transitions
$\psi (2S)\to\gamma \chi_{cJ}(1P)$
and 30\%-one for
$ \chi_{cJ}(1P)\to\gamma\psi (1S)$ (note that slightly smaller errors
were obtained in the overall fit of Ref. \cite{of}).

The wave function parameters $c_i^{(n)}$ determined in (\ref{21})
are presented in Table 3 for Solutions I and II.  Correspondingly, in
Figs. 6 and 7 we demonstrate the wave functions for $\psi (nS)$ and
$\eta_c$, $\chi_{c0}$, $\chi_{c1}$, $\chi_{c2}$.
Comparing the wave functions depicted in Figs. 6 and 7, one can clearly
see that Solutions I and II differ unsignificantly. We have carried out
our calculations with two variants of the interaction in order to
make clear that the description of $c\bar c$ system  does not require a
variety of the $t$-channel exchanges and  inclusion of all the
versions given by Eq. (15) would only result in the absence of
convergence in fitting procedure. In particular, the considered example
of two sorts of interactions demonstrate that the $c\bar c$ system does
not require instanton-induced forces, which were needed for treating
the mass splitting of $\pi,\eta,\eta'$ \cite{inst}.

\section{Determination of the $c\bar c $ component of the photon wave
function}

The vertex function of the transition $\gamma\to\cbc$ is represented
with the following formula:
\bea
\label{22}
G_{\gamma\to\cbc}(s)=\sum\limits_{n=1}^6 C_n
 G_{\psi(nS)}(s)
+\frac 1{1+\exp[(-\beta_\gamma (s-s_0)])}\ ,
\eea
where
 $ G_{\psi(nS)}(s)=
\Psi_{\psi(nS)}(s)
(s-M_{\psi(nS)}^2)$ and
$M_{\psi(nS)}$ and $\Psi_{\psi(nS)}(s)$ are given in Tables 2
and 3; $C_n$,  $\beta_\gamma$ and $s_0$ are the parameters to be
determined.

\subsection{Decay $\psi(nS)\to e^+e^-$}

Partial width for the decay $\psi(nS)\to e^+e^-$ reads:
\bea
\label{23}
\Gamma(\psi(nS)\to e^+e^-)=
\frac{\pi\alpha^2}{M_{\psi(nS)}^5}
\sqrt{\frac{M_{\psi(nS)}^2-4\mu_e^2}{M_{\psi(nS)}^2}}
\left(\frac 83 \mu_e^2+\frac 43 M_{\psi(nS)}^2\right)
\left |F_{\psi(nS)\to e^+e^-}(0)\right |^2,
\eea
where $\alpha=e^2/4\pi=1/137$,
$\mu_e$ is the electron mass, and $M_{\psi(nS)}$ is the charmonium mass.
The transition amplitude
$F_{\psi(nS)\to e^+e^-}(0)$, being determined by the process of Fig. 4,
see  \cite{1}, is equal to:
\bea
\label{24}
F_{\psi(nS)\to e^+e^-}(0)=\frac 23 \sqrt{N_c}
\int\limits_{4m_c^2}^\infty \frac {ds}{16\pi^2}\Psi_{\psi(nS)}(s)
G_{\gamma\to c\bar c}(s)
\sqrt{1-\frac{4m_c^2}s}\left(\frac 83 m_c^2+\frac 43 s\right).
\eea
The wave function in (\ref{24}) is normalized as follows:
\bea
\label{25}
1=\int\limits_{4m_c^2}^\infty \frac{ds}{16\pi^2}\Psi_{\psi(nS)}^2(s)
\sqrt{1-\frac{4m_c^2}s}\left(\frac 83 m_c^2+\frac 43 s\right)\ .
\eea
The coeficients  $c_i^{(n)}$ given in Table 3
construct the wave functions obeying this normalization constraint.

\subsection{Decay $\eta_c (nP)\to \gamma\gamma$}

Partial width for the decay $\eta_c\to\gamma\gamma$ reads:
\bea
\label{26}
\Gamma(\eta_c\to\gamma\gamma)=\frac\pi4
\alpha^2 M_{\eta_c}^3 \left |F_{\eta_c\to\gamma\gamma}(0)\right |^2\ .
\eea
The transition amplitude is determined by the processes of Figs. 1a,b;
it is equal to \cite{1,4,epja}:
\bea
\label{27}
F_{\eta_c(nP)\to\gamma\gamma}(0)=\frac 89 \sqrt{N_c}\,  m_c
\int\limits_{4m_c^2}^\infty \frac {ds}{2\pi^2}\Psi_{\eta_c(nP)}(s)
\Psi_{\gamma \to c\bar c}(s)
\ln\frac{\sqrt{s}+\sqrt{s-4m_c^2}}{\sqrt{s}-\sqrt{s-4m_c^2}},
\eea
where $N_c =3$.
Recall that $\Psi_{\gamma \to c\bar c}(s)=G_{\gamma \to c\bar c} (s)/s
$\ .

Normalization condition for pseudoscalar charmonium wave functions is as
follows:
\bea
\label{28}
1=\int\limits_{4m_c^2}^\infty
\frac{ds}{8\pi^2}\Psi_{\eta_c(nP)}^2(s) \sqrt{1-\frac{4m_c^2}s}\,s\ .
\eea
Coefficients presented in Table 3 give us $\Psi_{\eta_c(nP)}(s)$ obeying
(\ref{28}).

\subsection{Decay $\chi_{c0}(nP)\to \gamma\gamma$}

Partial width of the decay $\chi_{c0}\to\gamma\gamma$ is equal to
\bea
\label{29}
\Gamma(\chi_{c0}\to\gamma\gamma)=
\frac{\pi\alpha^2}{M_{\chi_{c0}}}
\left |F_{\chi_{c0}\to\gamma\gamma}(0)\right |^2\ ,
\eea
with the quark transition amplitude (Figs. 1a,b) determined as follows
\cite{epja,YFscalar}:
\bea
\label{30}
F_{\chi_{c0(nP)}\to\gamma\gamma}(0)=\frac 89 \sqrt{N_c}\, m_c
\int\limits_{4m_c^2}^\infty \frac {ds}{4\pi^2}\Psi_{\chi_{c0(nP)}}(s)
\Psi_{\gamma \to c\bar c}(s)
\left(\sqrt{s(s-4m_c^2)}
-2m_c^2\ln\frac{\sqrt{s}+\sqrt{s-4m_c^2}}{\sqrt{s}-\sqrt{s-4m_c^2}}
\right ) .
\eea
Normalization condition for scalar charmonium wave function reads:
\bea
\label{31}
1=\int\limits_{4m_c^2}^\infty \frac{ds}{8\pi^2}\Psi_{\chi_{c0(nP)}}^2(s)
\sqrt{1-\frac{4m_c^2}s}(s-4m^2_c)m^2_c\ .
\eea

\subsection{Decay $\chi_{c2}(nP)\to \gamma\gamma$}

Partial width $\chi_{c2}\to\gamma\gamma$ is defined by two transition
amplitudes:
\bea
\label{32}
\Gamma(\chi_{c2}\to\gamma\gamma)=\frac
45\frac{\pi\alpha^2}{M_{\chi_{c2}}} \left (\frac 16
\left |F_{\chi_{c2}\to\gamma\gamma}^{(0)}(0)\right|^2 +
\left|F_{\chi_{c2}\to\gamma\gamma}^{(2)}(0)\right|^2\right)\ ,
 \eea
which are determined by the processes of Figs. 1a,b \cite{epja,YFtensor}
 and   for the $P$-wave quark-antiquark states they read:
equal  to:
\bea
\label{33}
F_{\chi_{c2}(nP)\to\gamma\gamma}^{(H)}(0)=\frac 89 \sqrt{N_c}
\int\limits_{4m_c^2}^\infty \frac {ds}{16\pi^2}\Psi_{\chi_{c2}(nP)}(s)
\Psi_{\gamma \to c\bar c}(s)
 I^{(H)}(s)\ .
\eea
Here
\bea
I^{(0)}(s)=-2\sqrt{s\left(s-4m_c^2\right)}
\left(12m_c^2+s\right)
+4m_c^2\left(4m_c^2+3s\right)
\ln\frac{s+\sqrt{s\left(s-4m_c^2\right)}}
        {s-\sqrt{s\left(s-4m_c^2\right)}}\ ,
\label{34}
\eea
$$
\nonumber
I^{(2)}(s)=\frac{4\sqrt{s\left(s-4m_c^2\right)}}{3}
\left(5m_c^2+s\right)
-4m_c^2\left(2m_c^2+s\right)
\ln\frac{s+\sqrt{s\left(s-4m_c^2\right)}}
        {s-\sqrt{s\left(s-4m_c^2\right)}}\ .
$$
Normalization condition for the $P$-wave tensor $c\bar c$ system it is:
\bea
\label{36}
1=\int\limits_{4m_c^2}^\infty \frac{ds}{16\pi^2}\Psi_{\chi_{c2}(nP)}^2(s)
\frac 8{15}\sqrt{1-\frac{4m_c^2}s}(8m_c^2+3s)(s-4m_c^2)\ .
\eea

\subsection{The results of the fit}

By fitting the reactions involving the transition $\gamma\to c\bar c$,
we have determined the parameters $C_n,\beta_\gamma,s_0$
defined in (\ref{22}). For Solutions I and II, they are as follows (in
GeV units):
 \bea {\rm Solution\, I}&\qquad &{\rm Solution\,
II}\nonumber\\
    C_1=   -4.945& \qquad  & C_1=-4.995 \nonumber\\
    C_2=     -2.893&\qquad  &C_2=  -2.897 \nonumber\\
    C_3=     -2.191&\qquad  &C_3=  -2.179 \nonumber\\
    C_4=     -2.300&\qquad  &C_4=  -2.260 \nonumber\\
    C_5=     -4.264&\qquad  &C_5=  -4.368 \nonumber\\
    C_6=     -0.690&\qquad  &C_6=  -0.479 \nonumber\\
b_\gamma =     0.14&\qquad  &b_\gamma =  0.15 \nonumber\\
s_0 =     11.9&\qquad  &s_0 =  22.5
\label{37}
\eea
Experimental values of partial widths included into fitting procedure
versus those obtained in the fitting procedure are shown in
Table 4. There are also predictions made for the two-photon decays of
the first radial excitation states: $\eta_c(3594)$, $\chi_0(3847)$,
$\chi_2(3947)$.

Let us note that the decay $\chi_{c2}(3556)\to \gamma\gamma$ was not
included into the fit because of controversity of the data. In the
reaction $p\bar p\to \gamma\gamma$, the value $\Gamma(\chi_2(3556)\to
\gamma\gamma)=0.32\pm 0.080\pm 0.055$ keV was obtained \cite{E760},
while in direct measurements,  such as $e^+e^-$ annihilation,
 the width is much
larger:
$ 1.02\pm 0.40\pm 0.17\;{\rm keV}$ \cite{L3}\ ,
$1.76\pm 0.47\pm 0.40\;{\rm keV}$ \cite{OPAL}\  ,
$1.08\pm 0.30\pm 0.26 \;{\rm keV}$ \cite{CLEO}\  .
The compilation \cite{PDG} provides the value
close to  that of \cite{E760}. The value found in our fit agrees with
data reported by \cite{L3,OPAL,CLEO} and contradicts
\cite{E760}.

\begin{table}
\caption{Calculated partial widths for $nS$ and $nP$ states
{\it versus} experimental data
(bold mass numbers stand for the predicted states)}
\begin{center}
\begin{tabular}{|c|c|c|}
\hline
Decay                          & $\Gamma$ (keV)&$\Gamma_{exp}$ (keV)\\
\hline
$1S:\qquad J/\psi(3096)      \to e^+e^-$ & 5.444 (I)   & 5.40 $\pm$ 0.22
\\
                               & 5.598  (II)       &                    \\
\hline
$2S:\qquad   \psi(3686)      \to e^+e^-$ & 2.151 (I)        & 2.14
$\pm$ 0.21    \\
                               & 2.210  (II)       &                    \\
\hline
$3S:\qquad   \psi(4040)      \to e^+e^-$ & 0.756 (I)
& 0.75 $\pm$ 0.15    \\
                               & 0.778  (II)       &                    \\
\hline
$4S:\qquad   \psi(4415)      \to e^+e^-$ & 0.462 (I)        & 0.47
$\pm$ 0.10    \\
                               & 0.498  (II)       &                    \\
\hline
\hline
$1S:\qquad \eta_c(2979)      \to \ggam  $ & 6.979 (I)
& 7.0  $\pm$ 1.0
\\
                                & 6.946 (II)       &                    \\
\hline
$2S:\qquad \eta_c(3594)      \to \ggam  $ & 1.968 (I)       & --- \\
                                & 1.034 (II)       &                    \\
\hline
\hline
$1P:\qquad \chi_{c0}(3415)      \to\ggam$ & 2.572 (I)       & 2.6
$\pm$ 0.5 \\
                                & 2.440 (II)       &                    \\
\hline
$2P:\qquad \chi_{c0}({\bf 3849})\to\ggam$ & 1.159 (I)       & ---
\\
$3P:\qquad \chi_{c0}({\bf 3845})\to\ggam$ & 1.021 (II)       & \\
\hline \hline
$1P:\qquad \chi_{c2}(3556)      \to\ggam$ & 1.195 (I)       & ---             \\
                                & 1.155 (II)       &                    \\
\hline
$2P:\qquad \chi_{c2}({\bf 3950})\to\ggam$ & 2.051 (I)       & ---
\\ $3P:\qquad \chi_{c2}({\bf 3943})\to\ggam$ & 1.934 (II)       & \\ \hline
\end{tabular} \end{center} \end{table}

\section{Conclusion}

We have carried out the calculations of radiative transitions, where
the $c\bar c$ system participates, and compared the results with the
experiment. The results are given in Table 5. In general, there is a
 good description of the data. Still, one should point to a
disagreement for the following two cases: $\psi(2S)\to
\chi_{c1}(1P)\gamma$ and $J/\psi\to \eta_c(1S)\gamma$.

The calculation of partial width $\psi(2S)\to \chi_{c1}(1P)\gamma$
provides us with a value twice as large as given in \cite{c1-1,c1-2}.
Such a disagreement may be related to either presumably much higher
experimental error \cite{c1-1,c1-2} or a specific
behaviour of the wave function of $\chi_{c1}(1P)$, that was not
accounted for in \cite{4}.

Another discrepancy has been observed for the width of the transition
$J/\psi\to \eta_c(1S)\gamma$. This is an M1-transition, it is defined
by the magnetic moment of the $c$-quark. One possibility to reduce the
calculated value of $J/\psi\to \eta_c(1S)\gamma$ consists in the
increase of the $c$-quark mass, another one is to include into
consideration anomalous magnetic moment of $c$-quark. The hypothesis of
the existence of  anomalous magnetic moment at light quarks was
suggested rather long ago in connection with the description of the
decay $\omega\to \pi^0\gamma$ \cite{Az}, see also discussion
in \cite{Ger}.

\begin{table}
\caption{Comparison of experimental data (in keV units)
with our results and calculations of other groups.}
\begin{center}
{\footnotesize
\begin{tabular}{|l|c|c|c|c|c|c|c|}
\hline
Decay & Data & results & LS(F)\cite{Linde} & LS(C)\cite{Linde}
&RM(S)\cite{Resag}&RM(V)\cite{Resag}&NR\cite{NR}\\
\hline
$\psi (2S)\to\chi_{c0}(1P)\gamma$  & 26$\pm$ 4  & 22& 31--47   & 26--31
& 31 & 32 &19.4\\
 $\psi(2S)\to\chi_{c1}(1P)\gamma$  & 25$\pm$ 4  & 59&
58--49   & 63--50  & 36 & 48 &34.8\\
$\psi(2S)\to\chi_{c2}(1P)\gamma$
& 20$\pm$ 4  & 19& 48--47   & 51--49  & 60 & 35 &29.3\\
\hline
$\psi(2S)\to\eta_{c }(1S)\gamma$  &0.8$\pm$0.2 &0.4& 11--10   & 10--13
& 6  &1.3 &4.47\\
\hline $\chi_{c0}(1P)\to J/\psi(1S)\gamma$&165$\pm$50
&169&130--96   &143--110 &140 &119 &147 \\
$\chi_{c1}(1P)\to
J/\psi (1S)\gamma$&295$\pm$90 &389&390--399  &426--434 &250 &230 &287 \\
$\chi_{c2}(1P)\to J/\psi(1S)\gamma$&390$\pm$120&229&218--195  &240--218
&270
 &347 &393 \\
\hline
$J/\psi (1S)\to\eta_{c }(1S)\gamma$
&1.1$\pm$0.3&4.1&1.7--1.3&1.7--1.4&3.35&2.66&1.21\\
\hline \hline
$J/\psi (1S)\to e^+ e^-       $ &5.4 $\pm$0.22&5.44&5.26      &5.26
 &8.05&9.21&12.2\\
$\psi(2S)\to e^+ e^-       $
&2.12$\pm$0.12&2.15&2.8--2.5&2.9--2.7&4.30&5.87&4.63\\
$\psi(3S)\to e^+
e^-       $ &0.75$\pm$0.15&0.76&2.0--1.6&2.1--1.8&3.05&4.81&3.20\\
$\psi(4S)\to e^+ e^-       $
&0.47$\pm$0.10&0.46&1.4--1.0&1.6--1.3&2.16&3.95&2.41\\
\hline
$\eta_{c }(1S)\to\gamma\gamma$ &7.0 $\pm$0.9 &6.98&6.2--6.3  &6.2--6.5  &4.2 &3.8 &19.1\\
$\chi_{c0}(1P)\to\gamma\gamma$ &2.6 $\pm$0.5 &2.57&1.6--1.8  &1.5--1.6  &--  &--  &--  \\
\hline
$\chi_{c2}(1P)\to\gamma\gamma$
&1.02$\pm$0.40$\pm$0.17\cite{L3}&1.17&0.3--0.4&0.3--0.4& --&--&--\\
                               &1.76$\pm$0.47$\pm$0.40\cite{OPAL}& & & & & & \\
                               &1.08$\pm$0.30$\pm$0.26\cite{CLEO}& & & & & & \\
                               &0.33$\pm$0.08$\pm$0.06\cite{E760}& & &
                               & & & \\
\hline
\end{tabular}
}
\end{center}
\end{table}

In Table 5, the values of partial width are presented obtained by the
other authors.

In \cite{Linde}, the ideology of treating the $c\bar c$ system
is similar to ours: the charmonium masses were fitted as well as
the widths of
radiative transitions. The results obtained in \cite{Linde}
depend on a chosen gauge for gluon exchange interaction
--- we demonstrate the results obtained  both
for Feynman (F) and Coulomb (C) gauges; different approaches used
in \cite{Linde}  are
reflected in the allowed accuracy intervals given in Table 5.

In \cite{Resag} the $c\bar c$ system was studied in terms of scalar (S)
and vector (V) confinement forces --
both variants are presented in Table 5. For the comparison we give in
Table 5 the results
obtained in the nonrelativistic approach to the $c\bar c$
system.

Both in relativistic \cite{Linde,Resag} and nonrelativistic
\cite{NR} approaches there is rather large discrepancy between the
data and calculated values of $\psi(nS)\to e^+e^-$ (in \cite{Linde} the
width of the transition $J/\psi\to e^+e^-$ was not calculated but fixed).
In our opinion, the reason is that in all
above-mentioned papers, soft interaction of quarks
was not accounted for --- we mean the processes shown in Fig. 3b,c.
In fact the necessity of taking into consideration the low-energy quark
interaction, that is, the vector meson dominance in the transitions
$q\bar q\to V\to \gamma$, was understood decades ago
 but till now this procedure has not become commonly
accepted even for light quarks: see, for example,
\cite{deWitt,Xiao}.

\section*{Acknowledgments}

We thank A.V. Anisovich, Y.I. Azimov, G.S. Danilov, I.T. Dyatlov,
L.N. Lipatov, V.Y. Petrov, H.R. Petry and M.G. Ryskin
for useful discussions.

This work was supported by the Russian Foundation for Basic Research,
project no. 04-02-17091.

\newpage
\begin{figure}
\centerline{\epsfig{file=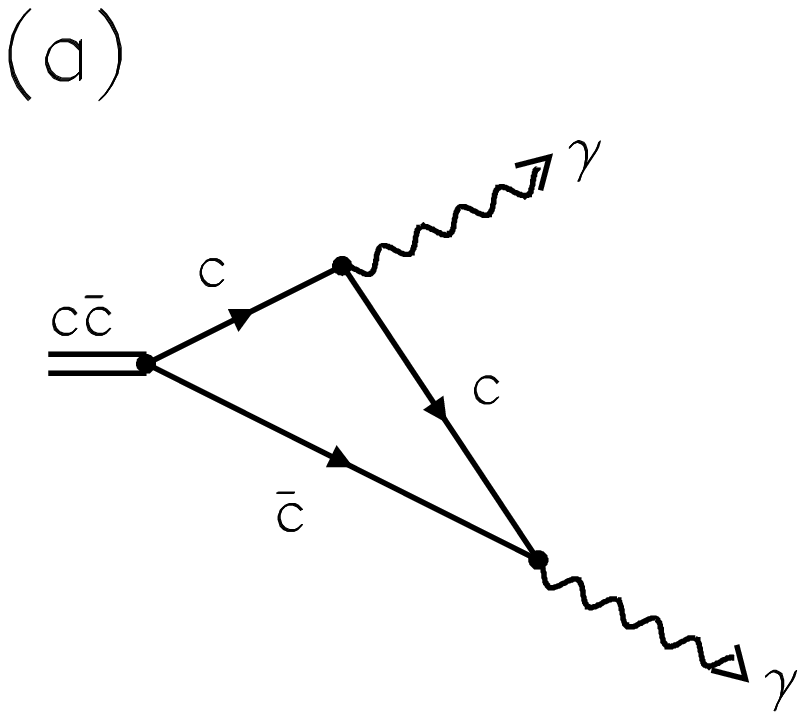,width=6cm}\hspace{0cm}
            \epsfig{file=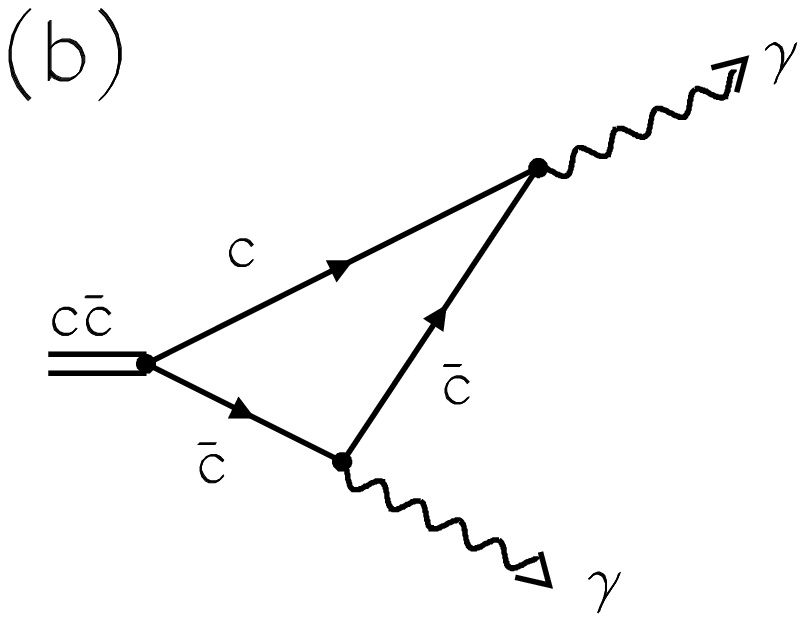,width=6cm}\hspace{0cm}
            \epsfig{file=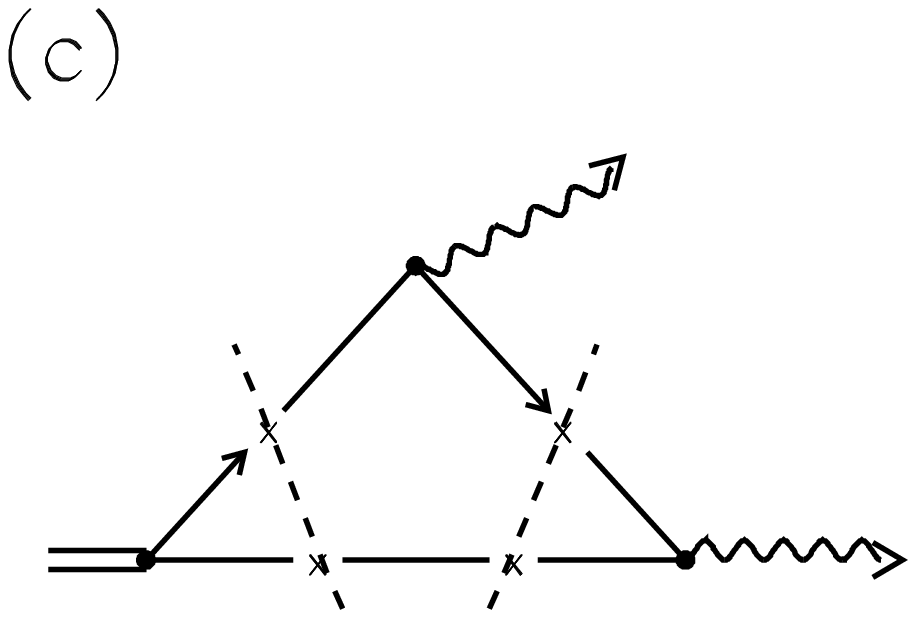,width=6cm}}
\vspace{0.5cm}
\caption{a,b) Diagrams for the two-photon decay of $c\bar c$ state,
c) Cuttings in the spectral integral representation.}
\end{figure}

\begin{figure}
\centerline{\epsfig{file=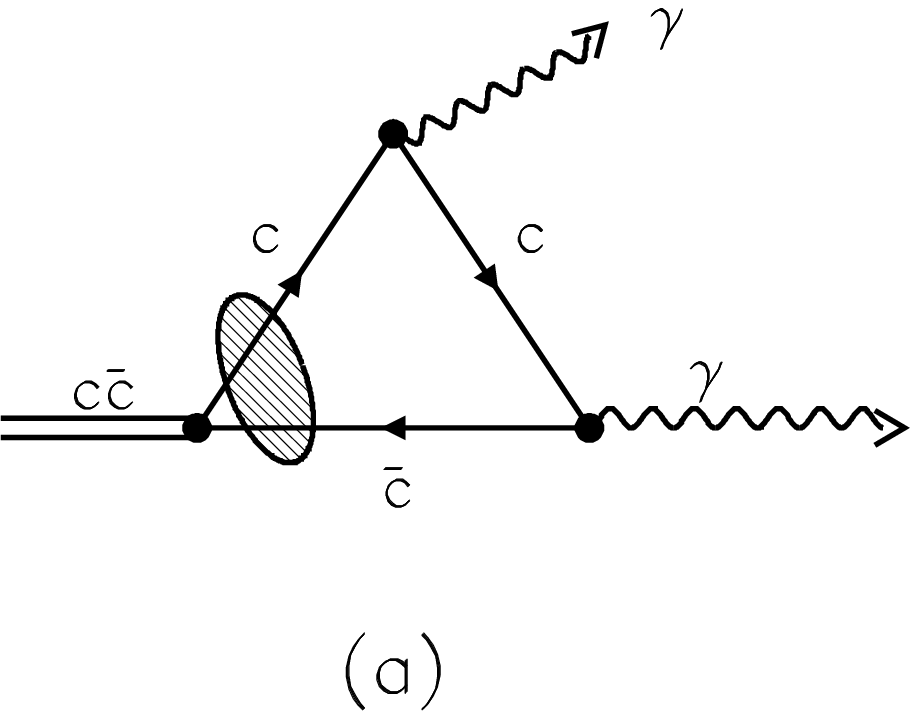,width=6cm}\hspace{1cm}
            \epsfig{file=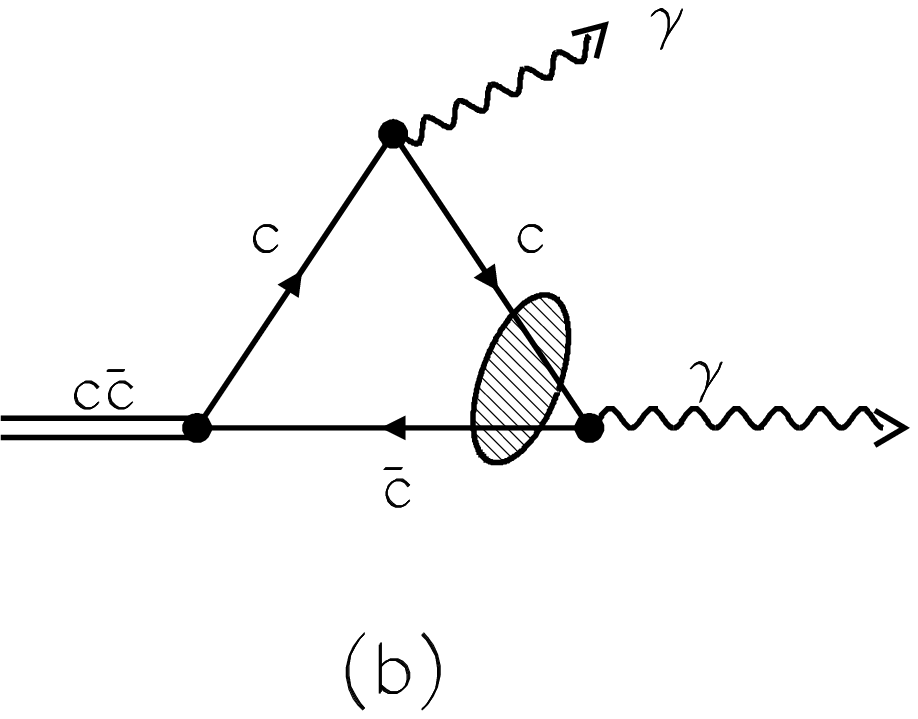,width=6cm}}
\vspace{0.5cm}
\caption{Initial (a) and final (b) state interactions of quarks in the
decay diagrams.}
\end{figure}

\begin{figure}
\centerline{\epsfig{file=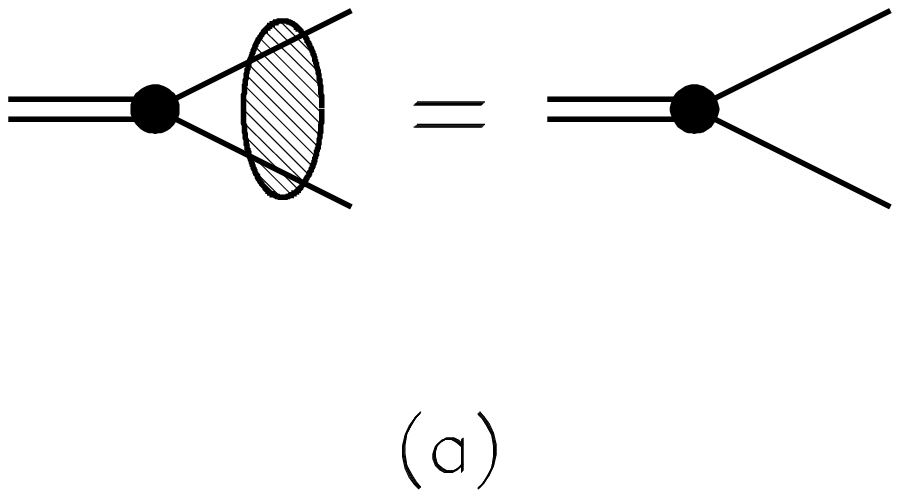,width=6cm}\hspace{1cm}
            \epsfig{file=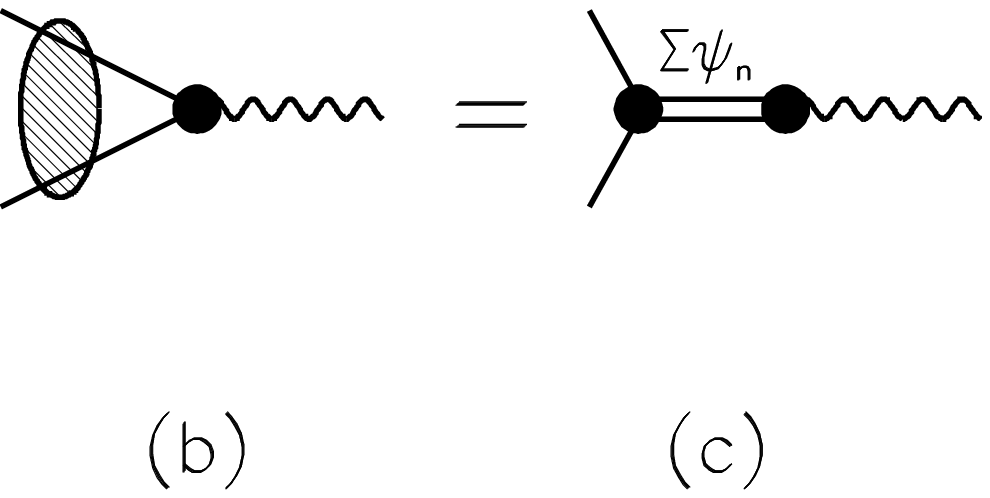,width=6cm}}
\vspace{0.5cm}
\caption{a) Graphical representation of the
Bethe-Salpeter equation for the $c\bar c$ vertex; c,d) interaction of
quarks in the vertex $c\bar c\to\gamma $ and its
approximation by the sum of transitions
$c\bar c \to \Sigma \psi (nS) \to \gamma $ .}
\end{figure}

\begin{figure}
\centerline{\epsfig{file=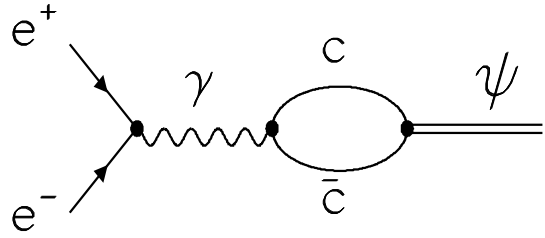,width=9cm}}
\vspace{0.5cm}
\caption{Quark transition diagram for the process $e^+e^-\to\psi$ .}
\end{figure}

\begin{figure}
\centerline{\epsfig{file=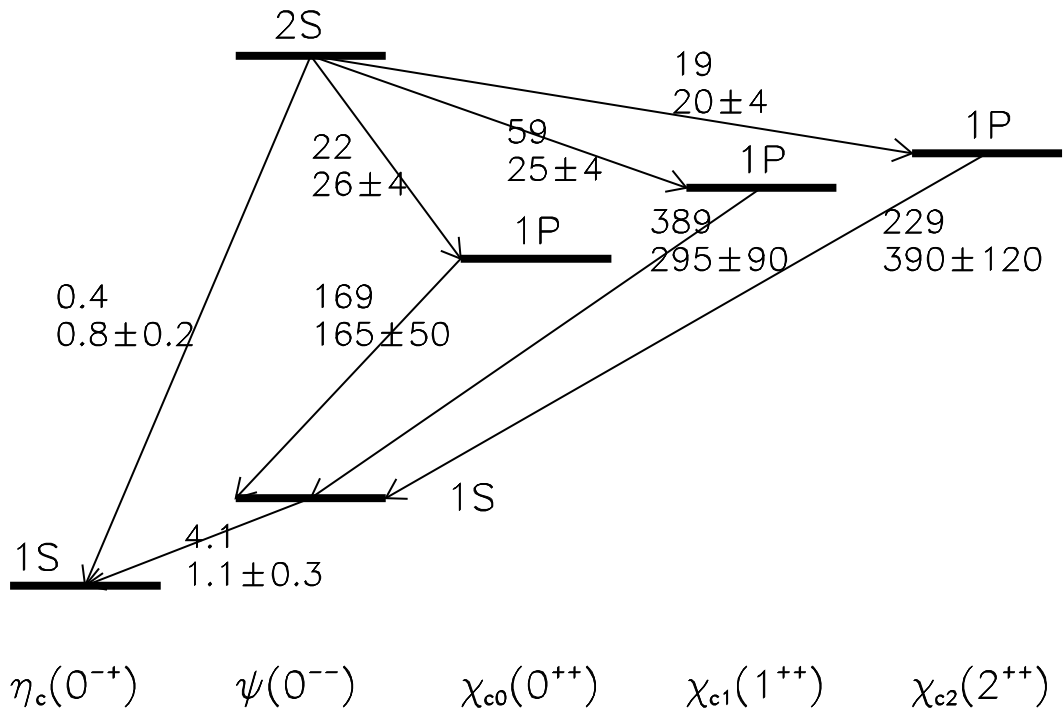,width=15cm}}
\vspace{1.0cm}
\caption{Radiative decays of the charmonium systems, which
were taken into account in the fit [4]. The
calculated decay partial widths
are shown in the keV units
(upper numbers) together with experimental data and errors
accepted in the fit (lower numbers).}
\end{figure}

\newpage
\begin{figure}
\centerline{\epsfig{file=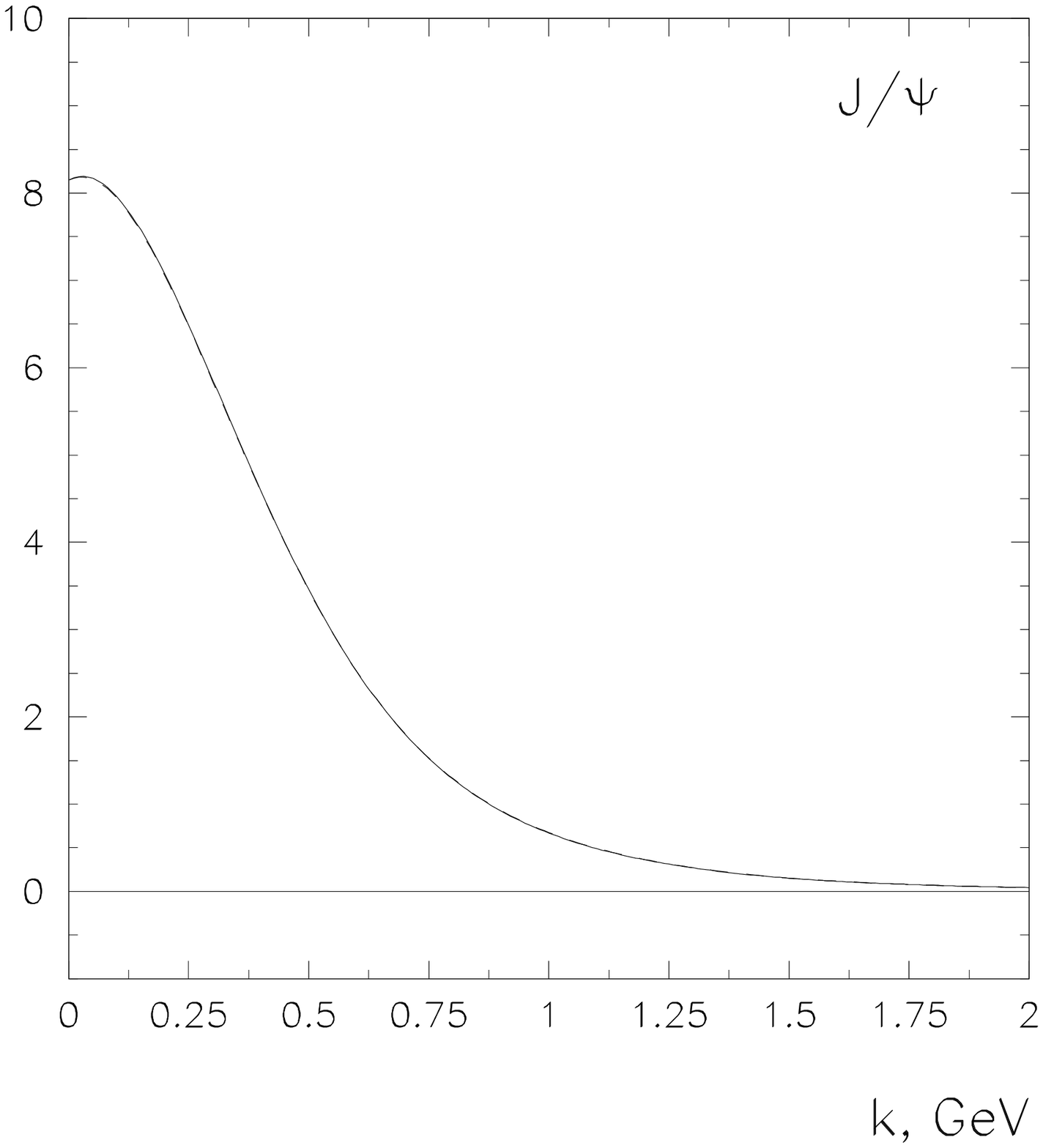,width=7cm}
            \epsfig{file=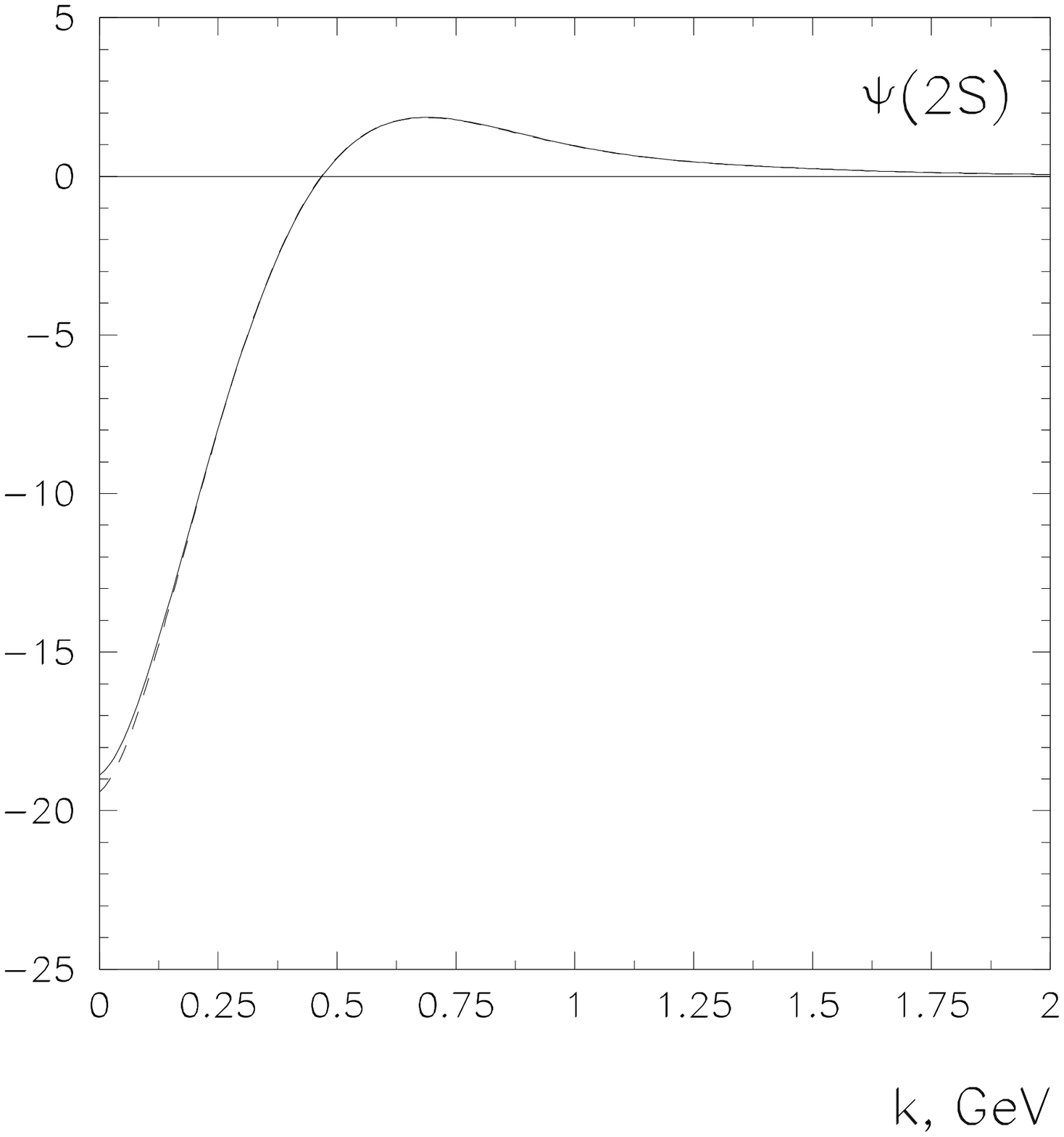,width=7cm}}
\vspace{-0.5cm}
\centerline{\epsfig{file=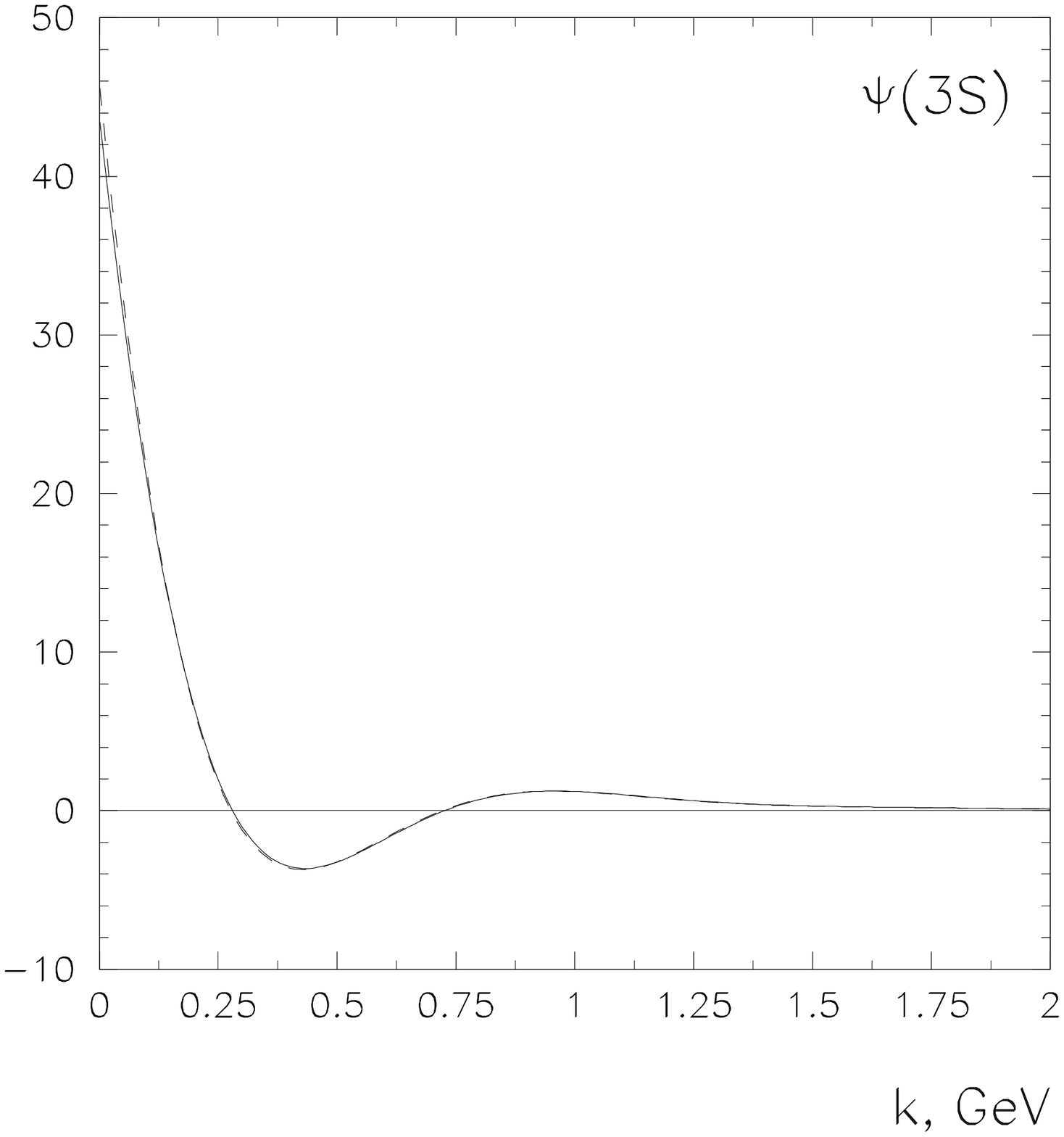,width=7cm}
            \epsfig{file=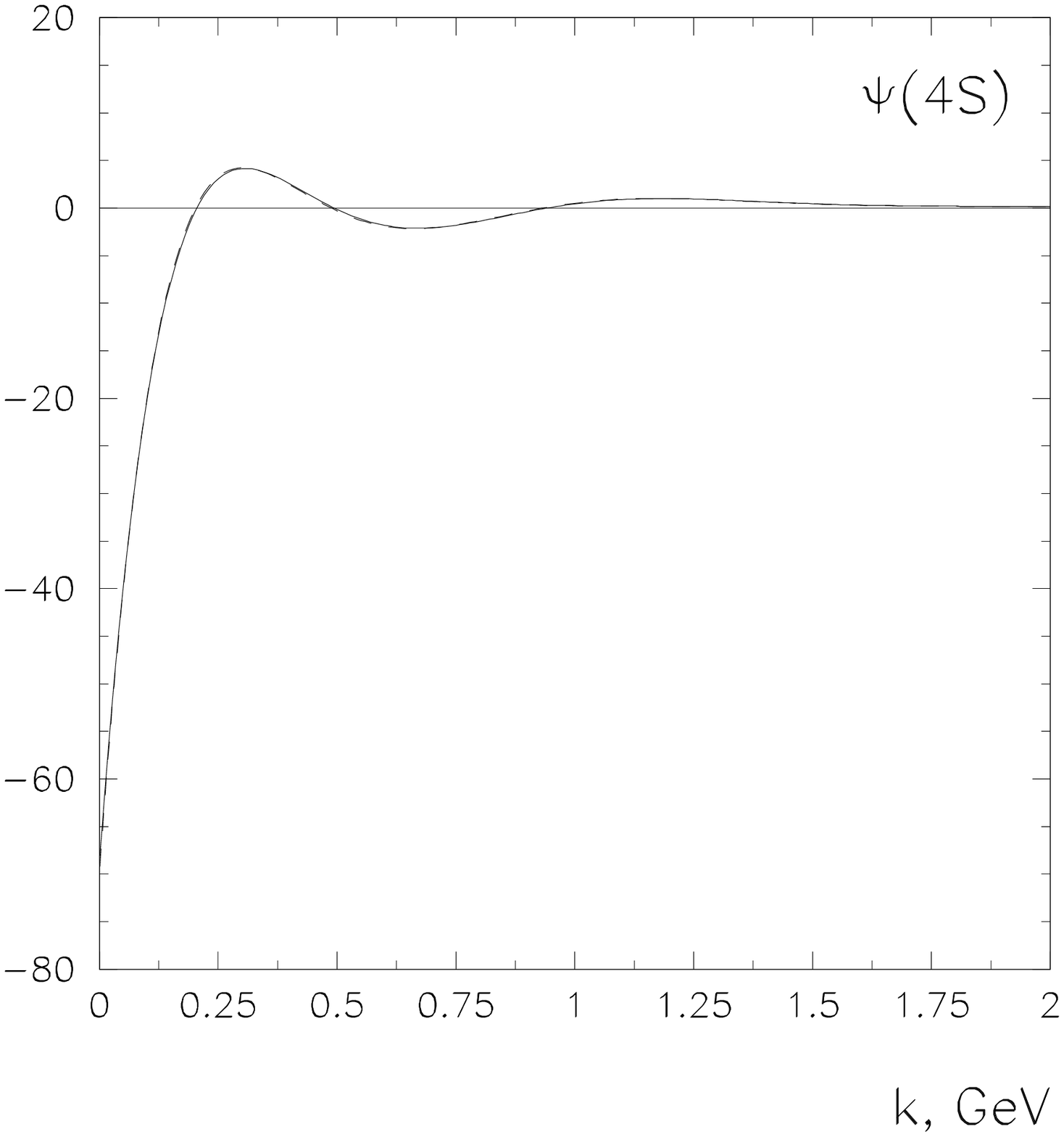,width=7cm}}
\vspace{-0.5cm}
\centerline{\epsfig{file=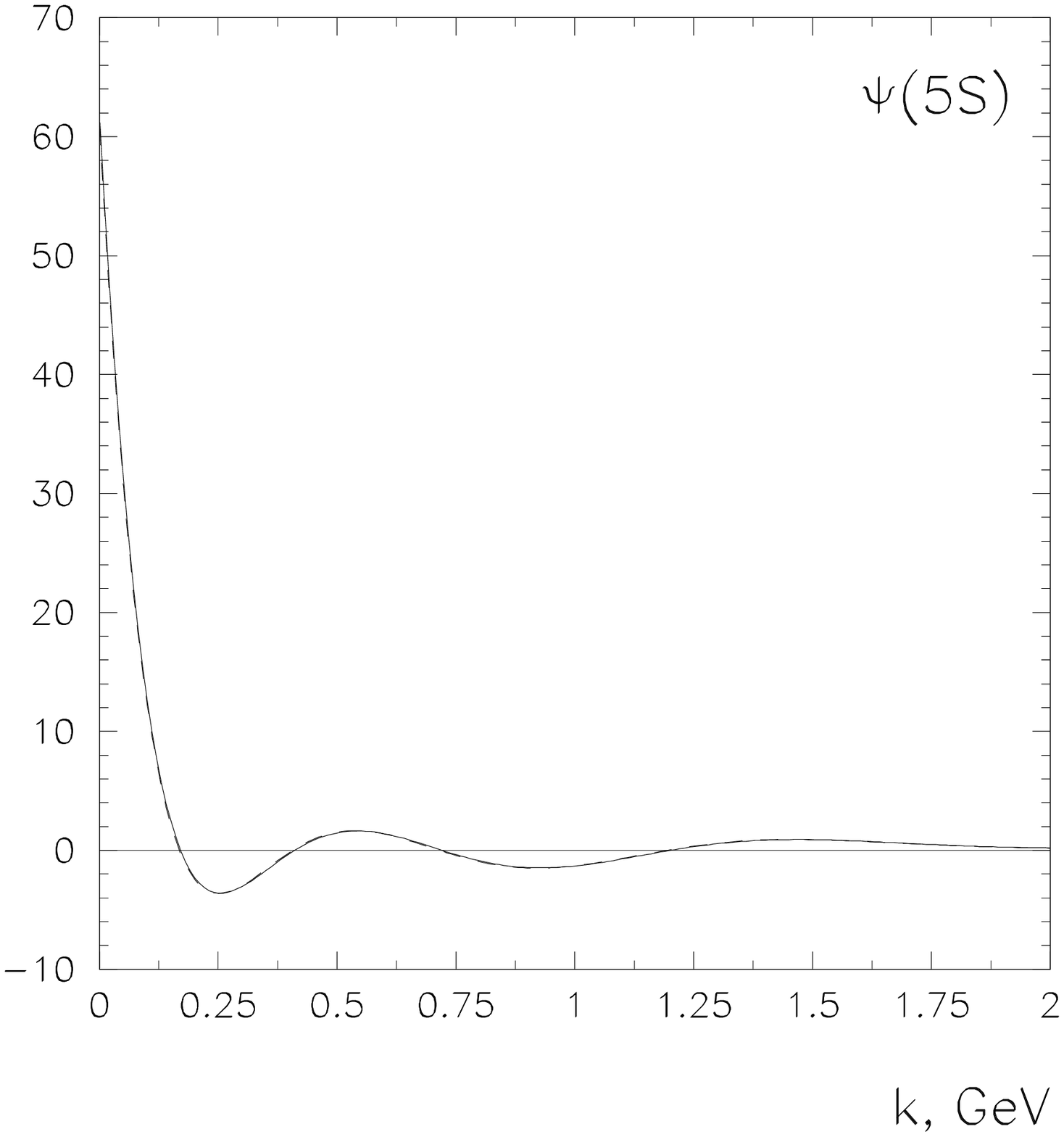,width=7cm}
            \epsfig{file=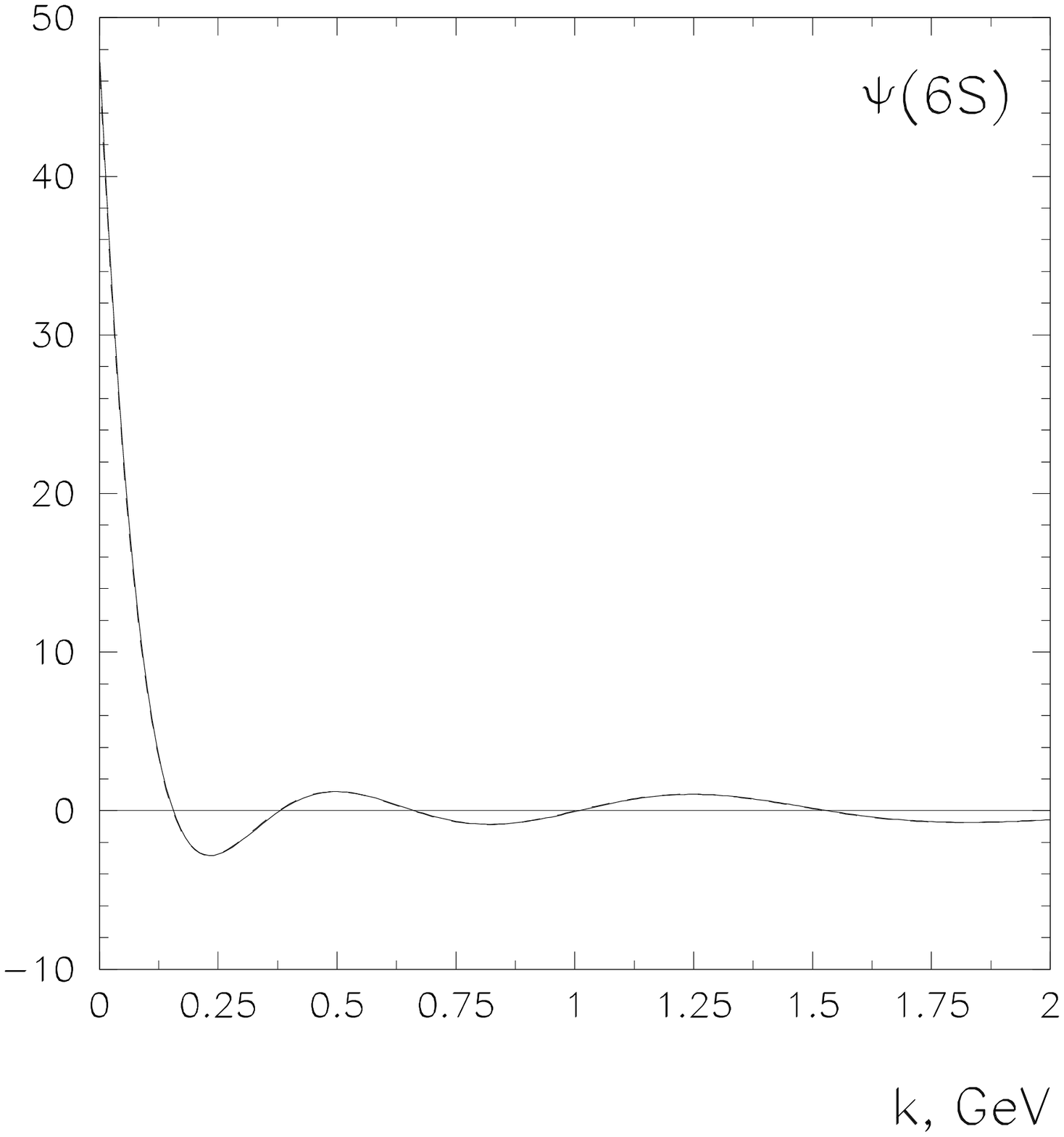,width=7cm}}
\caption{The $c\bar c$ wave functions for $\psi (nS)$. Solid and
 dashed lines stand for Solutions I and II.}
 \end{figure}

\newpage
\begin{figure}
\centerline{\epsfig{file=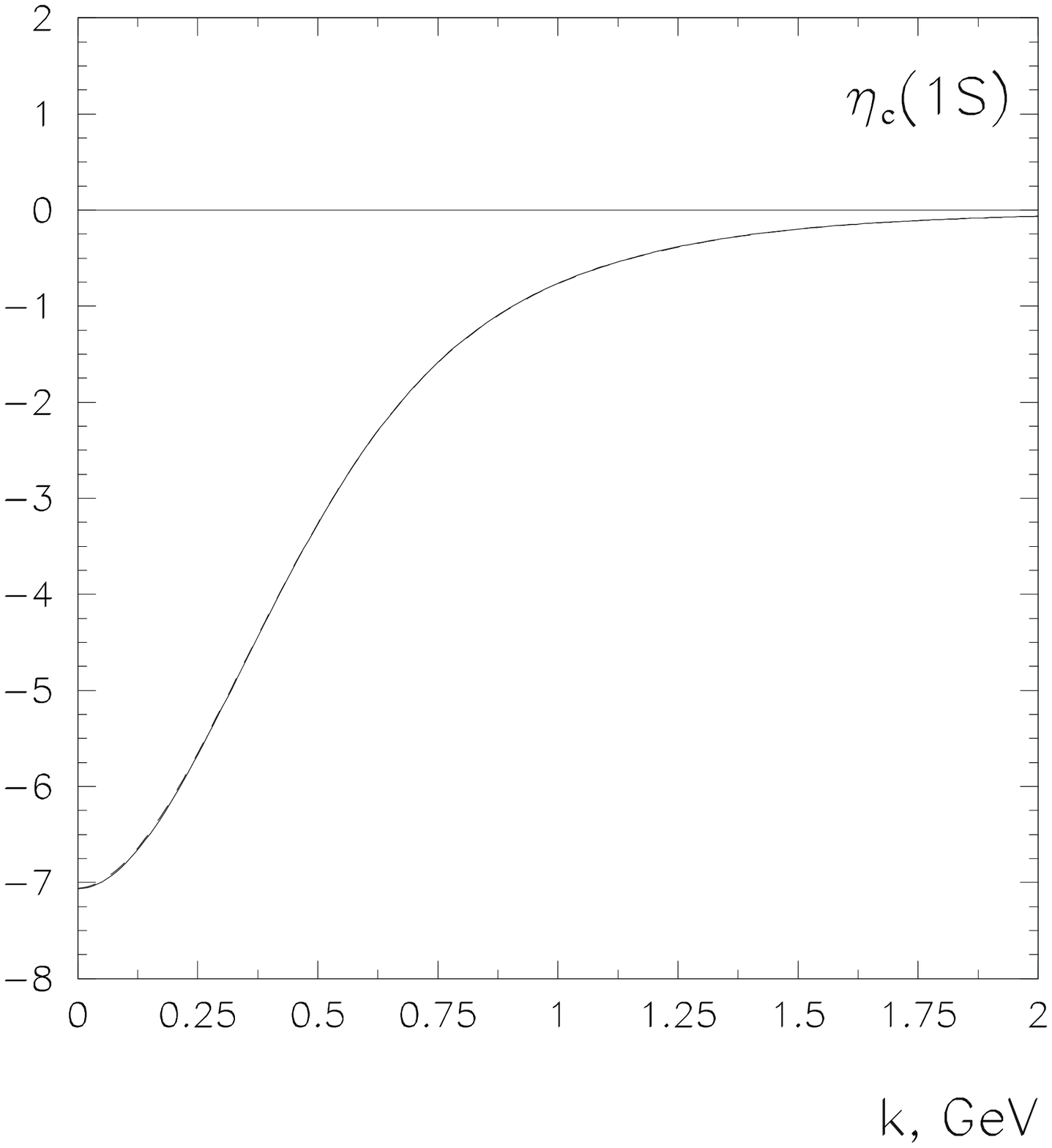,width=7cm}
            \epsfig{file=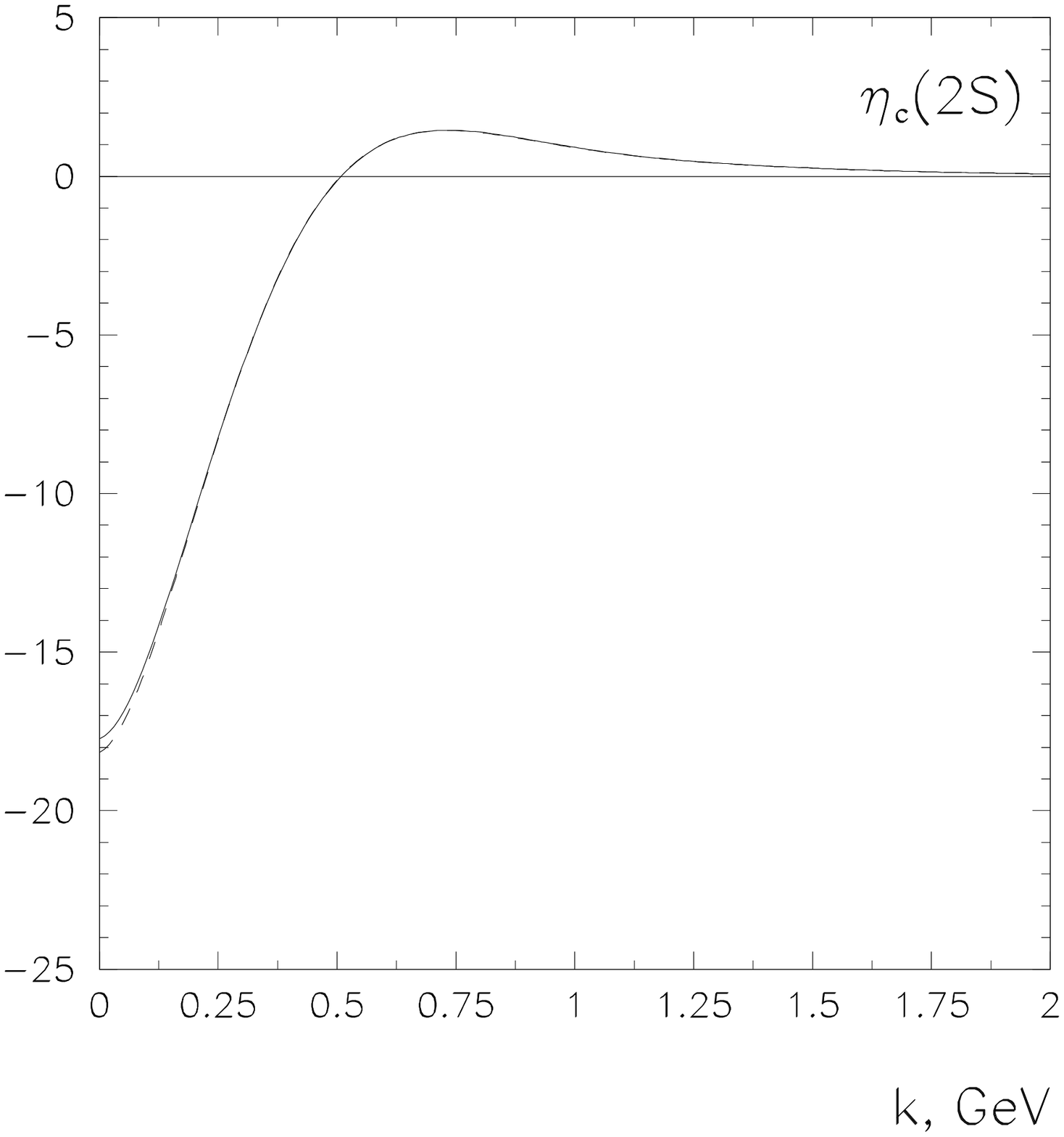,width=7cm}}
\vspace{-0.5cm}
\centerline{\epsfig{file=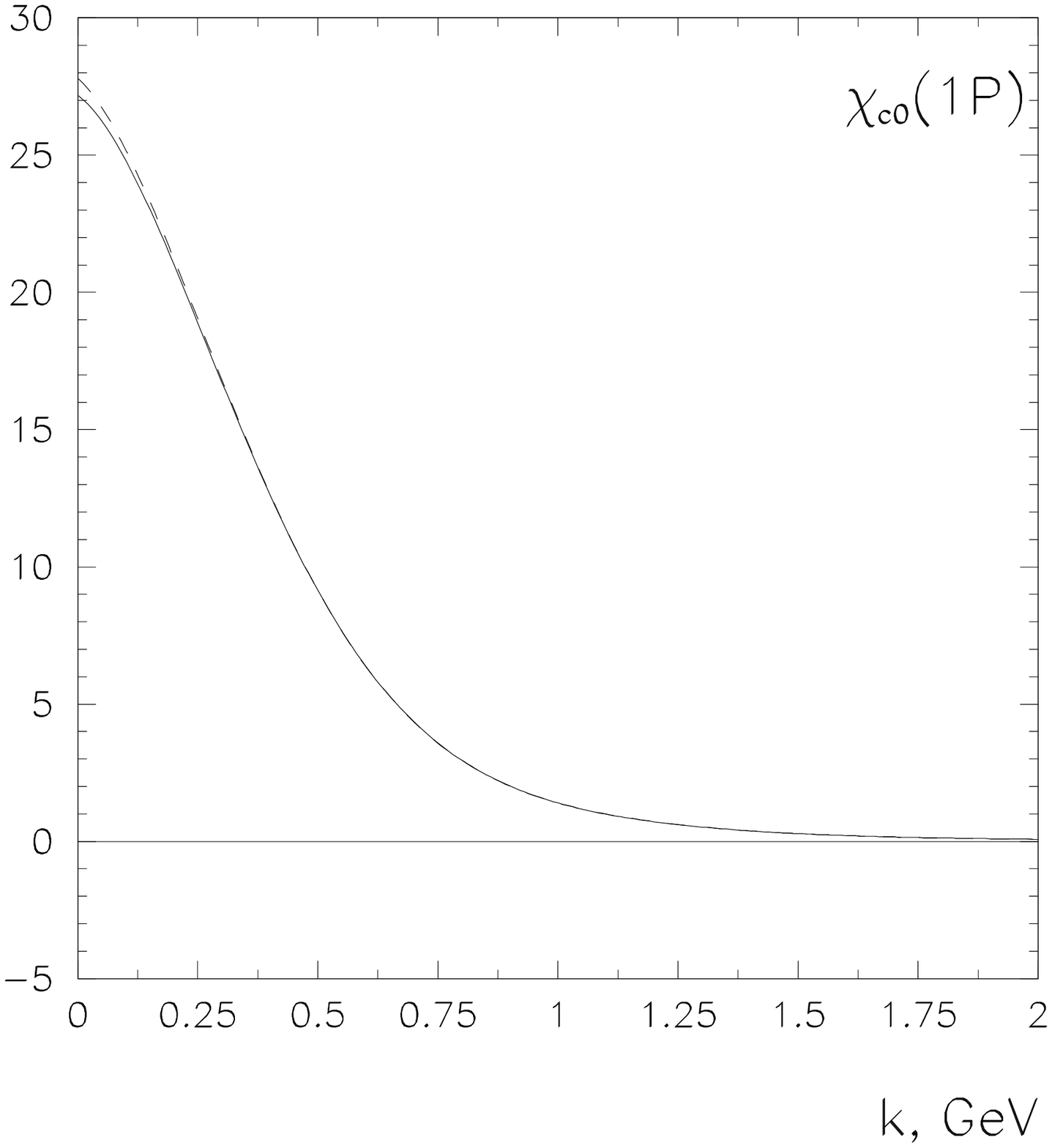,width=7cm}
            \epsfig{file=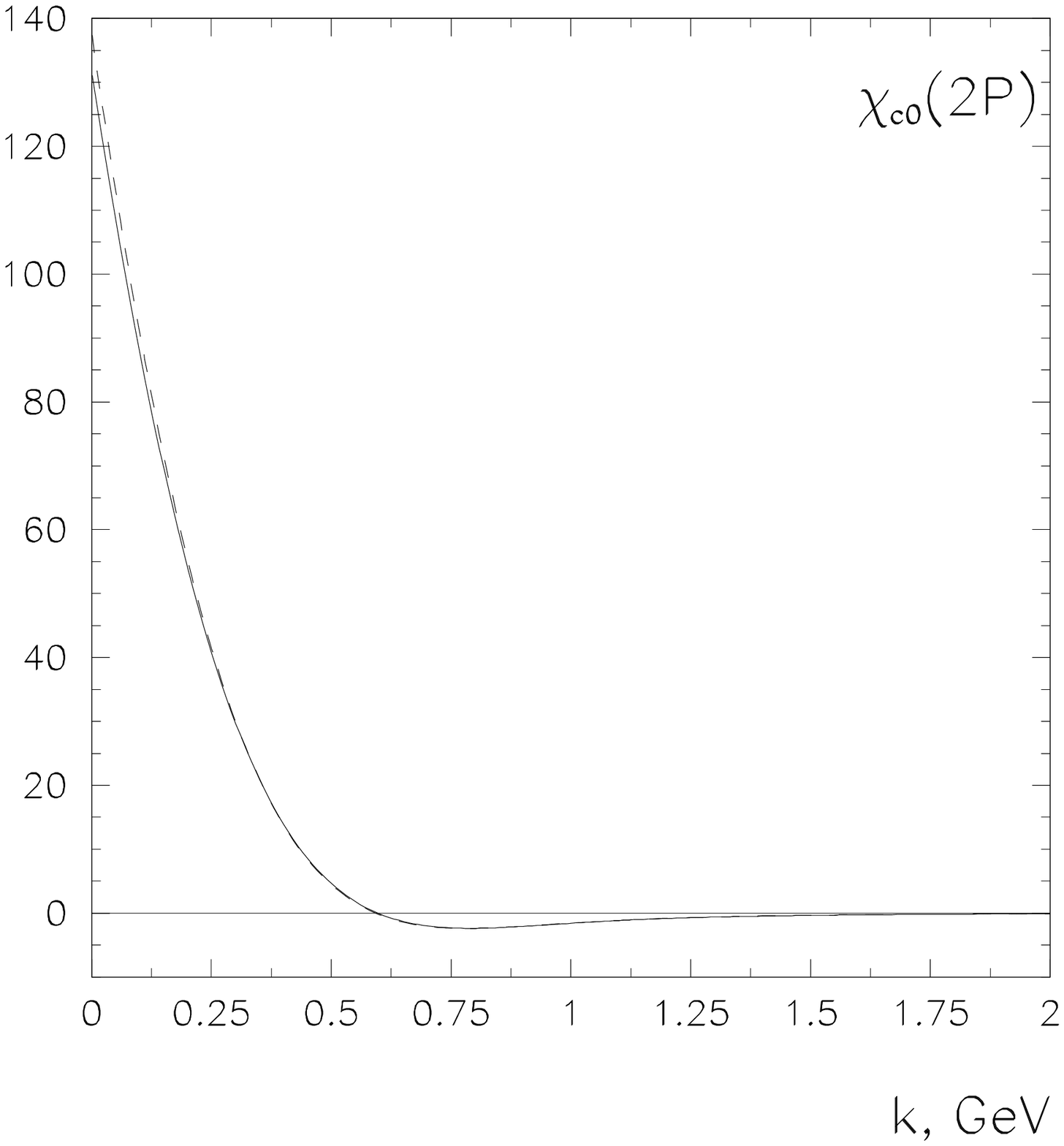,width=7cm}}
\vspace{-0.5cm}
\centerline{\epsfig{file=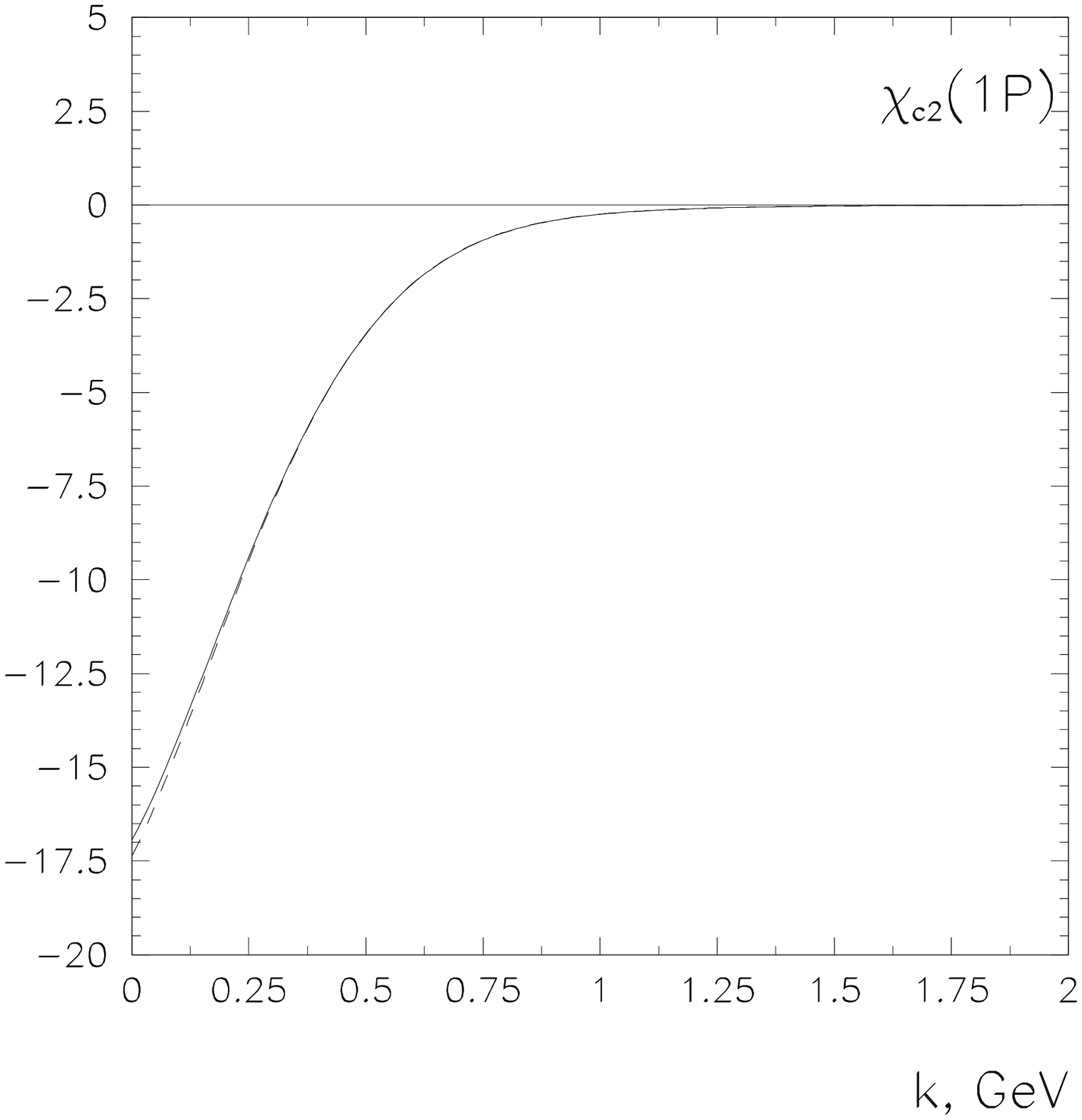,width=7cm}
            \epsfig{file=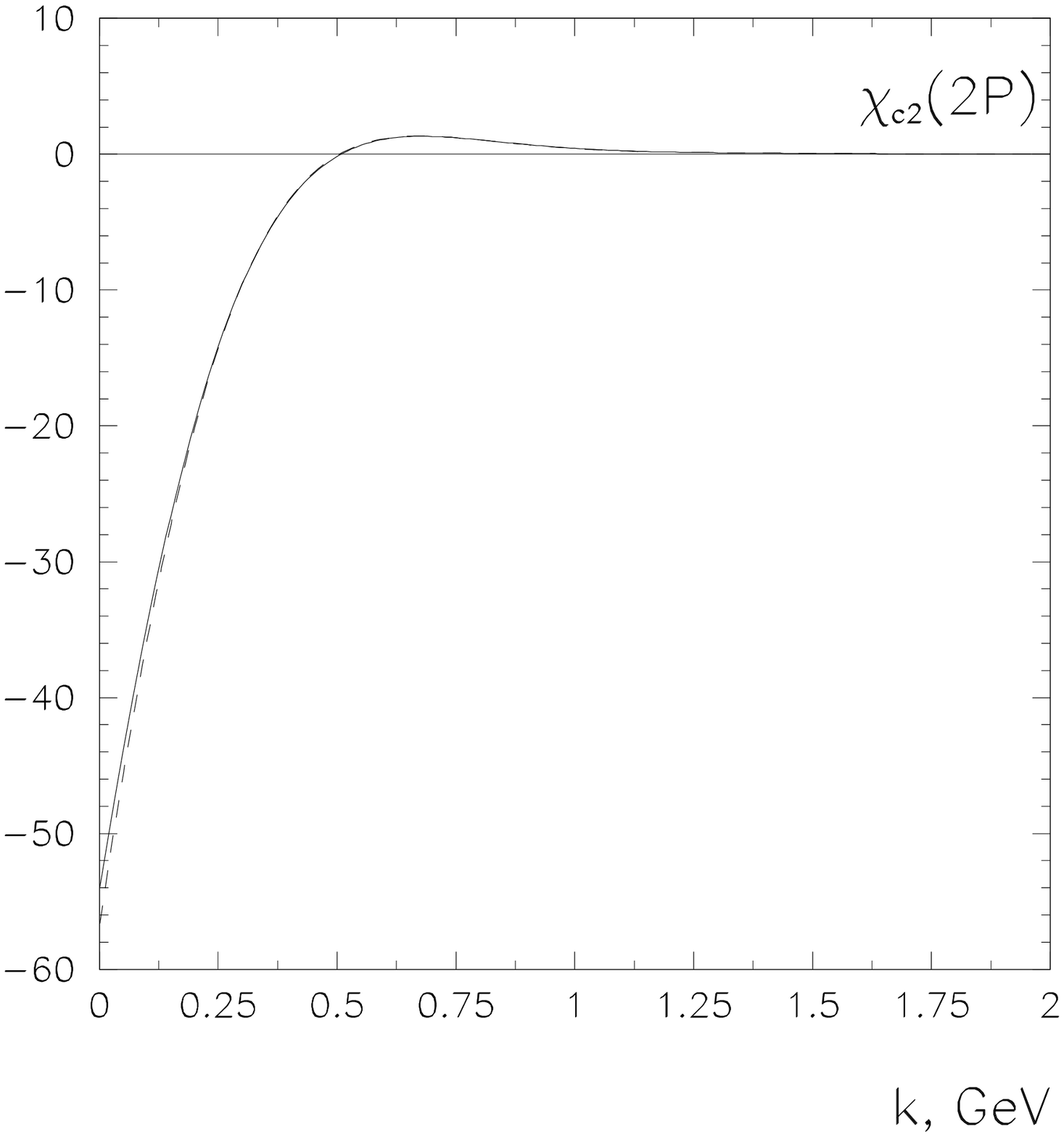,width=7cm}}
\caption{The $c\bar c$ wave functions for $\eta_c (nS)$, $\chi_{c0}(nP)$
and $\chi_{c2}(nP)$. Solid and   dashed lines correspond to Solutions I
and II.}
 \end{figure}

\begin{figure}
\centerline{\epsfig{file=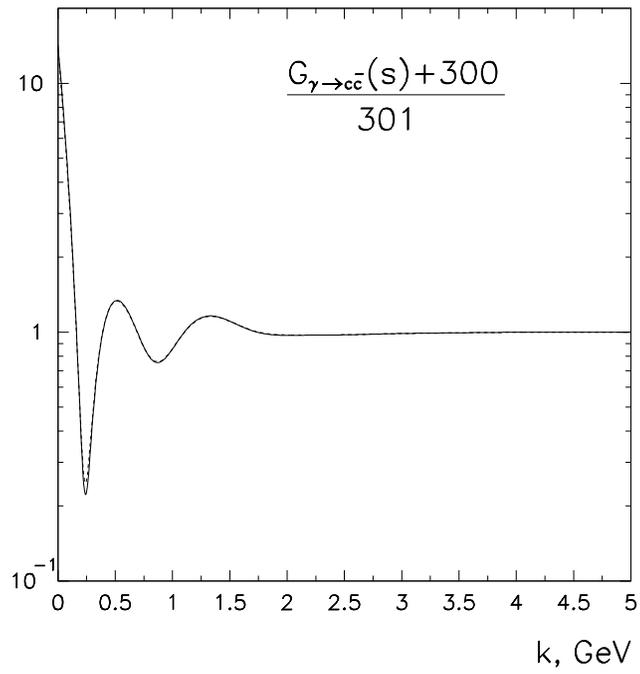,width=10cm}}
\caption{Vertex $\gamma\to\cbc$ ; solid and dashed lines correspond
to Solutions I and II.}
\end{figure}

\end{document}